\documentclass[iop]{emulateapj}
\usepackage{psfig,amsfonts,amsmath,graphicx,natbib,apjfonts,subfigure}
\def\nod{\nodata}
\def\swift{{\it Swift}}

\def\ein{1}
\def\ari{2}
\def\col{3}
\def\har{4}

\shorttitle{Radio Emission from Short GRBs}
\shortauthors{Fong et al.}

\begin{document}

\title{Radio Constraints on Long-Lived Magnetar Remnants in Short Gamma-Ray Bursts}

\author{W.~Fong\altaffilmark{\ein}$^{,}$\altaffilmark{\ari}, B.~D.~Metzger\altaffilmark{\col}, E.~Berger\altaffilmark{\har}, \& F.~{\"O}zel\altaffilmark{\ari} }

\altaffiltext{\ein}{Einstein Fellow}
\altaffiltext{\ari}{Steward Observatory, University of Arizona, 933 N. Cherry Ave, Tucson, AZ 85721}
\altaffiltext{\col}{Columbia Astrophysics Laboratory, Columbia University, New York, NY 10027}
\altaffiltext{\har}{Harvard-Smithsonian Center for Astrophysics, 60 Garden Street, Cambridge, MA 02138}

\begin{abstract}
The merger of a neutron star binary may result in the formation of a rapidly-spinning magnetar. The magnetar can potentially survive for seconds or longer as a supramassive neutron star before collapsing to a black hole if, indeed, it collapses at all.  During this process, a fraction of the magnetar's rotational energy of $\sim10^{53}$ erg is transferred via magnetic spin-down to the surrounding ejecta. The resulting interaction between the ejecta and the surrounding circumburst medium powers a $\gtrsim$ year-long synchrotron radio transient.  We present a search for radio emission with the Very Large Array following nine short-duration gamma-ray bursts (GRBs) at rest-frame times of $\approx1.3-7.6$~years after the bursts, focusing on those events which exhibit early-time excess X-ray emission that may signify the presence of magnetars.  We place upper limits of $\lesssim 18-32\,\mu$Jy on the 6.0 GHz radio emission, corresponding to spectral luminosities of $\lesssim (0.05-8.3) \times 10^{39}$~erg~s$^{-1}$. Comparing these limits to the predicted radio emission from a long-lived remnant and incorporating measurements of the circumburst densities from broad-band modeling of short GRB afterglows, we rule out a stable magnetar with an energy of $10^{53}$~erg for half of the events in our sample.  A supramassive remnant that injects a lower rotational energy of $10^{52}$~erg is ruled out for a single event, GRB\,050724A.  This study represents the deepest and most extensive search for long-term radio emission following short GRBs to date, and thus the most stringent limits placed on the physical properties of magnetars associated with short GRBs from radio observations.
\end{abstract}

\section{Introduction}

The merger of two neutron stars (NSs) in a compact binary can result in the formation of a massive NS remnant, which is generally assumed to collapse subsequently to a black hole. Accretion onto the black hole then powers a relativistic transient, a short-duration gamma-ray burst (GRB; \citealt{npp92,rj99a,ajm+05,rgb+11,bfc13,tlf+13,ber14,rlp+16}), with a prompt gamma-ray emission duration of $\lesssim 2$~sec. One of the biggest uncertainties in this canonical picture is how long the NS remnant survives prior to collapse.  This depends on the mass of the final remnant and the highly uncertain Equation of State (EoS) of dense nuclear matter \citep{obg10,lhr+14,fbr+15,ltb+15,opg+16}.

A massive NS remnant, which is supported against gravity exclusively by its differential rotation, is known as a {\it hypermassive} NS.  Somewhat less massive NSs, which can be supported even by their solid body rotation, are known as {\it supramassive}.  A hypermassive NS can survive for at most a few hundred milliseconds after the merger, before collapsing due to the loss of its differential rotation by internal electromagnetic torques and gravitational wave radiation (e.g.,~\citealt{st06}).  In contrast, supramassive remnants spin-down to the point of collapse through less efficient processes, such as magnetic dipole radiation, and hence can remain stable for $\gtrsim$seconds to minutes.  The discovery of NSs with masses $\approx 2~M_{\odot}$ \citep{dpr+10,afw+13} places a lower limit on the maximum NS mass, making it likely that the remnants produced in at least some NS mergers are supramassive (e.g.,~\citealt{opr+10}). The mergers of particularly low mass binaries may even produce {\it indefinitely stable} remnants, from which a black hole never forms (e.g., ~\citealt{gp13}). 

The high angular momentum of a merging binary guarantees that the NS remnant will be born rotating rapidly, with a spin period close to the break-up value of $\sim\!1$~ms.  The remnant may also acquire a strong magnetic field, $\gtrsim\!10^{14}-10^{15}$~G, as a result of shear-induced instabilities and dynamo activity \citep{dt92,uso92,pr06,zm13}.  Such a supramassive ``magnetar'' remnant possesses a reservoir of rotational energy up to $\approx 10^{53}$~erg \citep{mb14,mmk+15}, which is not available in cases where the NS promptly collapses to a black hole. Through its magnetic dipole spin-down, a magnetar remnant serves as a continuous power source, with the exact evolution of its spin-down luminosity dependent on the birth period and dipole magnetic field strength of the remnant \citep{zm01,mqt08,bmt+12,scr14,sc16}.

\tabletypesize{\small}
\begin{deluxetable*}{lcccccc}
\tablecolumns{7}
\tablewidth{0pc}
\tablecaption{Log of VLA 6.0 GHz Observations 
\label{tab:obs}}
\tablehead {
\colhead {GRB}                &
\colhead {$z$}           &
\colhead {UT Date}           &
\colhead {$\delta t_{\rm rest}$}          &
\colhead {$F_{\nu}$} & 
\colhead {$\nu L_{\nu}$} &
\colhead {X-ray behavior}              \\
\colhead {}                    &
\colhead {}                 &
\colhead {}              &
\colhead {(yr)}           &
\colhead {($\mu$Jy)}  &   
\colhead {(erg s$^{-1}$)} &  
\colhead {}                                         
}
\startdata
GRB\,050724A & 0.257 &   2015 Feb 22.472 &   7.629   & $<22.1$ & $2.7 \times 10^{38}$ & Extended emission \\
GRB\,051221A & 0.546 &   2015 Feb 22.718 &  5.936  & $<19.5$ & $1.4 \times 10^{39}$ & Plateau \\
GRB\,070724A & 0.457 &  2015 Feb 21.058  &   5.206  & $<19.1$ & $9.1 \times 10^{38}$ & Plateau \\
GRB\,080905A & 0.122 &   2015 Feb 23.723 &  5.769 &  $<22.2$ & $5.2 \times 10^{37}$ & Plateau \\
GRB\,090510 & 0.903 &   2015 Mar 6.625 &  3.062  & $<26.5$ & $6.6 \times 10^{39}$ & Extended emission \\
GRB\,090515 & 0.403 &  2015 Mar 2.427 &   4.135 &  $<22.7$ & $8.0 \times 10^{38}$ & Plateau \\
GRB\,100117A & 0.915 &  2015 Feb 27.674 &   2.671  & $<32.0$ & $8.3 \times 10^{39}$ & Plateau$^{a}$ \\
GRB\,101219A & 0.718 &  2015 Feb 24.011 &    2.437  & $<17.5$ & $2.5 \times 10^{39}$ & Plateau \\
GRB\,130603B & 0.356 &  2015 Mar 5.451 &   1.292  & $<20.6$ & $5.4 \times 10^{38}$ & Late-time excess
\enddata
\tablecomments{Upper limits correspond to $3\sigma$ confidence. \\
$^{a}$ The X-ray afterglow of GRB\,100117A also exhibited flaring activity \citep{mcg+11}. }
\end{deluxetable*}

\citet{mqt08} showed that the ongoing energy input from a long-lived magnetar is well-matched to a puzzling feature which distinguishes a subset of short GRBs: $\approx\!$ 1/4$-$1/2 of short bursts discovered with the \swift\ satellite \citep{ggg+04} have excess emission in their light curves when compared to the standard synchrotron model for afterglows.  Indeed, $\approx\!15$-$20\%$ of \swift\ short GRBs have prolonged X-ray activity for tens to hundreds of seconds following the bursts themselves (``extended emission''; \citealt{nb06,pmg+09}). Other events have a temporary flattening or ``plateau'' in the flux decline rate of their X-ray afterglows for $\approx 10^2-10^3$ seconds after the burst \citep{nkg+06}. Still others have late-time excess X-ray emission on timescales of $\sim$few days \citep{pmg+09,fbm+14}. Other than the anomalous X-ray behavior, there are no obvious differences between these bursts and the normal population of short GRBs in any of their host galaxy properties \citep{fbc+13}, suggesting an origin intrinsic to the burst central engine.

Long-lived magnetar remnants have been commonly invoked to explain the excess emission observed following short GRBs. Several studies have fit magnetar models to short GRBs with extended emission \citep{gow+13}, X-ray and optical plateaus \citep{rot+10,rom+13,gvo+15}, and late-time excess emission \citep{fyx+13,fbm+14}, resulting in inferred spin periods of $\approx 1-10$~ms and large magnetic fields of $\approx (2-40) \times 10^{15}$~G.  As an alternative to the magnetar model, other energy sources remain energetically viable: most notably, late-time ``fall-back'' accretion onto the remnant black hole \citep{ros07,pnj08,ctg11}. In order to substantiate the magnetar scenario for this subset of short GRBs, and also provide crucial insight on the NS EoS, it is necessary to test additional predictions of the magnetar model.
  
Synchrotron radio emission is expected from the interaction of the ejecta in a NS merger and the surrounding circumburst medium \citep{np11}, similar to a young supernova remnant.  \citet{mb14} pointed out that the radio brightness of this interaction would be significantly enhanced in the case of a supramassive or stable magnetar remnant, due to the additional energy imparted to the ejecta by the injected rotational energy, which can exceed that of the dynamical ejecta by three or four orders of magnitude.  Since there is substantial observational evidence linking short GRBs to NS mergers \citep{fb13,bfc13,tlf+13,ber14}, radio observations following short GRBs offer an independent way to explore the existence of long-lived (supramassive or stable) magnetar remnants. Using radio observations of seven short GRBs on $\sim 1$-$3$~yr timescales, \citet{mb14} placed constraints on the circumburst density of $\lesssim 0.1-1$~cm$^{-3}$ assuming an energy reservoir of $3 \times 10^{52}$~erg. Similarly, \citet{hhp+16} analyzed two bursts and placed limits of $\lesssim 0.001-5$~cm$^{-3}$ depending on the value of the assumed ejecta mass, and assumed the same energy of $3 \times 10^{52}$~erg.

Here, we present radio observations of nine short GRBs on rest-frame timescales of $\sim2-8$ years after the bursts, focusing on those events which exhibit excess X-ray emission at early times that may signify the presence of magnetars. This sample represents the largest and deepest survey for long-timescale radio emission of short GRBs to date, and provides a unique test of the magnetar model. We utilize this data set to constrain the presence of magnetars formed as a result of the mergers. In Section~\ref{sec:obs}, we outline the sample and radio observations. In Section~\ref{sec:model}, we describe the magnetar model and in Section~\ref{sec:ar}, we present the analysis and results, including the constraints on the magnetar rotational energies and environment circumburst densities. In Section~\ref{sec:disc}, we compare this work to previous studies of emission from magnetars in short GRBs, and we conclude in Section~\ref{sec:conc}.

\section{Observations}
\label{sec:obs}

\subsection{Sample}

We select all short GRBs with sub-arcsecond localization, spectroscopic redshifts and sky locations observable with the VLA. We further require that the events have pre-existing observations which indicate unusual behavior potentially attributed to a remnant magnetar: extended emission, X-ray plateau, or a late-time X-ray excess. These selection criteria limit our sample to nine events (Table~\ref{tab:obs}). In our sample, two events have extended emission, six have X-ray plateaus, and one has a late-time excess.

\subsection{VLA Observations}

We observed the positions of nine short GRBs with the Karl G. Jansky Very Large Array (VLA) from 2015 Feb 21 to Mar 6 UT (Table~\ref{tab:obs}; Program 15A-246). For each burst, we obtained 1 hr of observations in B configuration at a mean frequency of $6.0$~GHz (lower and upper side-bands
centered at $4.9$~GHz and $7.0$~GHz). We follow standard procedures in the Astronomical Image Processing System (AIPS; \citealt{gre03}) for data calibration and analysis, using 3C48 and 3C286 for flux calibration, and standard sources as part of the VLA calibrator manual\footnote{https://science.nrao.edu/facilities/vla/observing/callist} for gain calibration. We do not detect any source in or around the positions of the GRBs. To obtain $3\sigma$ upper limits on the flux density, $F_{\nu}$, we use AIPS/{\tt IMSTAT} on source-free regions surrounding the GRB positions. The $6.0$~GHz upper limits are listed in Table~\ref{tab:obs}. The limits span a range, $F_{\nu}\lesssim 18-32\,\mu$Jy with a median of $F_{\nu}\lesssim 22\,\mu$Jy.

\bigskip

\section{Magnetar Model}
\label{sec:model}

Simulations of binary NS mergers find the dynamical ejection of $\sim0.01~M_{\odot}$ of material \citep{rlt+99,hkk+13, kio+15,rgl+16}, while a comparable or greater amount of mass may be lost in outflows from the remnant accretion disk \citep{mpq09,fm13,jbp+15}.  In the case of a long-lived NS remnant, the disk wind ejecta mass can approach the total disk mass of $\approx 0.1~M_{\odot}$ \citep{mf14}.  A high ejecta mass of $\sim\,0.03-0.08 ~M_{\odot}$ was also inferred based on modeling the kilonova emission from the short GRB\,130603B \citep{bfc13,hkt+13,tlf+13}.

In the case of a long-lived magnetar, a significant fraction of the magnetar's rotational energy can be imparted to the dynamical and disk wind ejecta, accelerating it to mildly relativistic speeds \citep{mp14,mb14}. The deceleration of this fast material by its shock interaction with the circumburst medium produces synchrotron emission peaking at MHz to GHz frequencies \citep{np11,mb14,hp15}. The synchrotron model provides a mapping from the flux densities to physical parameters of the magnetar and circumburst environment: magnetar's rotational energy ($E$), ejecta mass ($M_{\rm ej}$), circumburst density ($n$), fractions of post-shock energy in radiating electrons ($\epsilon_e$) and magnetic fields ($\epsilon_B$), and the electron power-law distribution index ($p$) that describes the input distribution of electrons with $N(\gamma)\propto \gamma^{-p}$.

In the radio band, the synchrotron spectrum is characterized by two break frequencies, the maximum frequency ($\nu_m$) and the self-absorption frequency ($\nu_a$). Here, we assume that the observing frequency is greater than both of the break frequencies, such that $\nu_{\rm obs}>\nu_m,\nu_a$, as is generally satisfied at $\nu_{\rm obs}$ = 6~GHz \citep{np11}. In this regime, the observed flux peaks at a characteristic deceleration timescale, $t_{\rm dec}$, by which time the ejecta transfers most of its energy to the surrounding medium. This timescale is given by \citep{np11}

\begin{equation}
t_{\rm dec} \approx 300 E_{52}^{-1/2} M_{\rm ej,-2}^{5/6} n^{-1/3}\,\,\,{\rm days}
\label{eqn:tdec}
\end{equation}

\noindent where $E_{52}$ is the energy in units of $10^{52}$~erg, $M_{\rm ej,-2}$ is the ejecta mass in units of $0.01~M_{\odot}$ and $n$ is the density in cm$^{-3}$. The corresponding peak flux density is \citep{np11}

\begin{equation}
F_{\nu,{\rm obs,pk}} \approx C E_{52}^{\frac{5p-3}{4}} M_{\rm ej,-2}^{-\frac{5p-7}{4}} n^{\frac{p+1}{4}} \epsilon_{B,-1}^{\frac{p+1}{4}} \epsilon_{e,-1}^{p-1} d_{L,27}^{-2} \nu_{\rm obs, 6}^{-\frac{p-1}{2}}\,\,\,\mu{\rm Jy}
\label{eqn:fpk}
\end{equation}

\noindent where the microphysical parameters $\epsilon_{e,-1}$ and $\epsilon_{B,-1}$ are in units of 0.1, $d_{L,27}$ is the luminosity distance in units of $10^{27}$~cm, $\nu_{\rm obs, 6}$ is the observing frequency in units of 6.0~GHz, and $C$ is a normalization constant, $3 \times 10^{5} \times 1.1^{\frac{5p-7}{2}} \times 4.3^{-\frac{p-1}{2}}$. For $\nu_{\rm obs}>\nu_m,\nu_a$, the observed flux evolves as

\begin{equation}
F_{\nu,{\rm obs}} =  \begin{cases} 
      F_{\nu,{\rm obs,pk}} \left(\dfrac{t}{t_{\rm dec}} \right)^{3} & t<t_{\rm dec} \\
      F_{\nu,{\rm obs,pk}} \left(\dfrac{t}{t_{\rm dec}} \right)^{-\frac{15p-21}{10}} & t\geq t_{\rm dec} \\
   \end{cases}
\label{eqn:fobs}
\end{equation}

\noindent where $t$ is the rest-frame time after the burst in days.

\begin{figure*}
\begin{minipage}[c]{\textwidth}
\tabcolsep0.0in
\includegraphics*[width=0.5\textwidth,clip=]{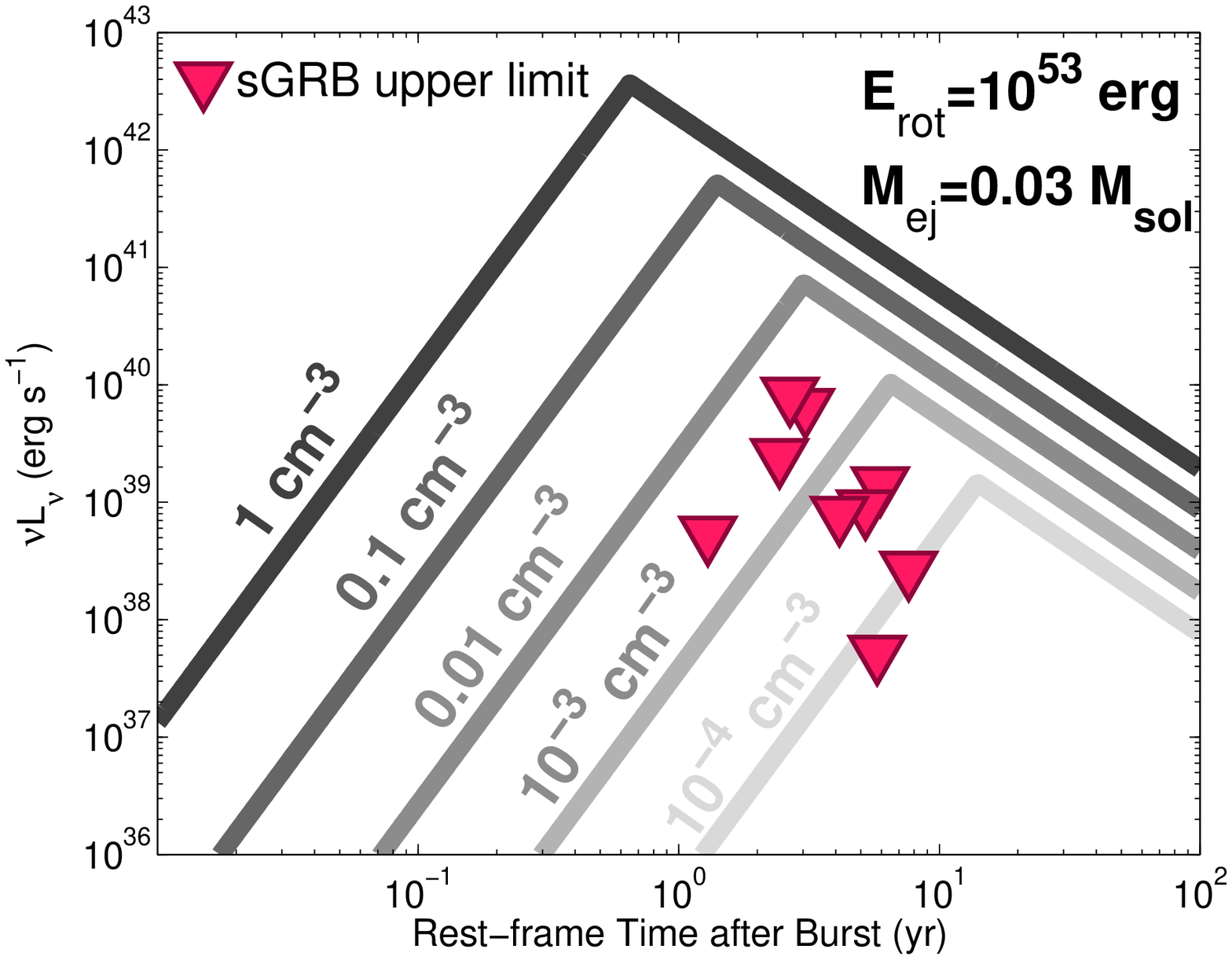}
\includegraphics*[width=0.5\textwidth,clip=]{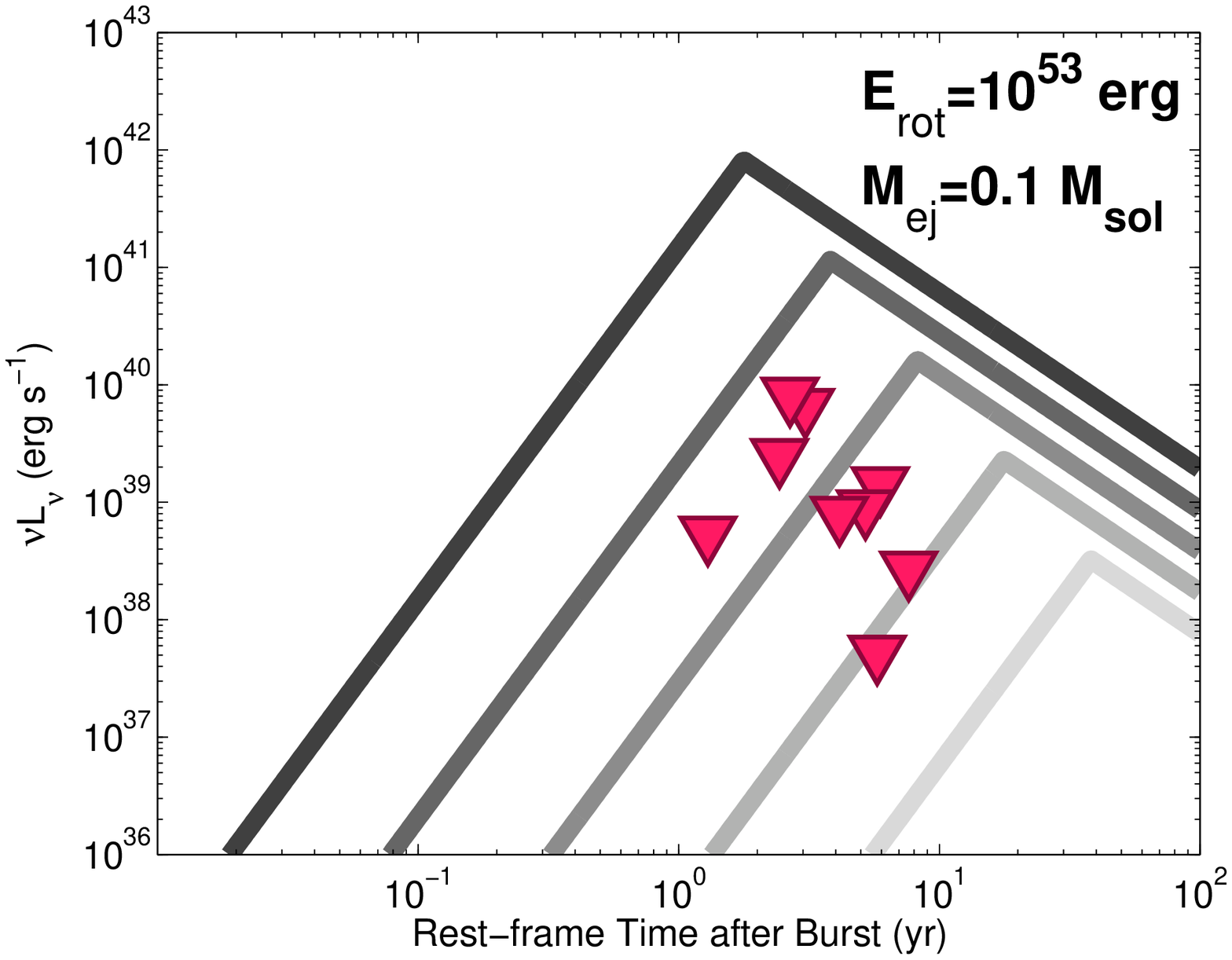} \\
\includegraphics*[width=0.5\textwidth,clip=]{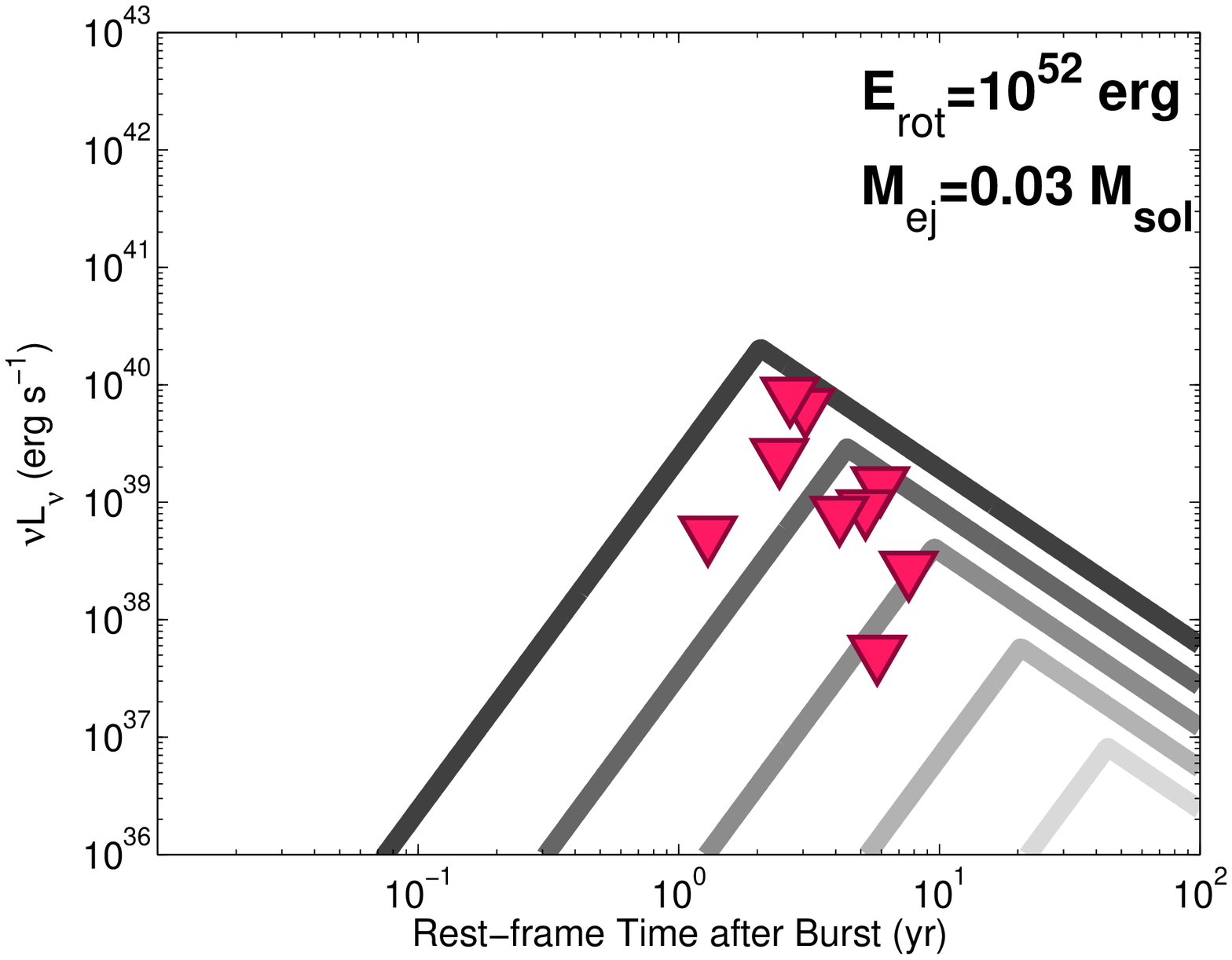}
\includegraphics*[width=0.5\textwidth,clip=]{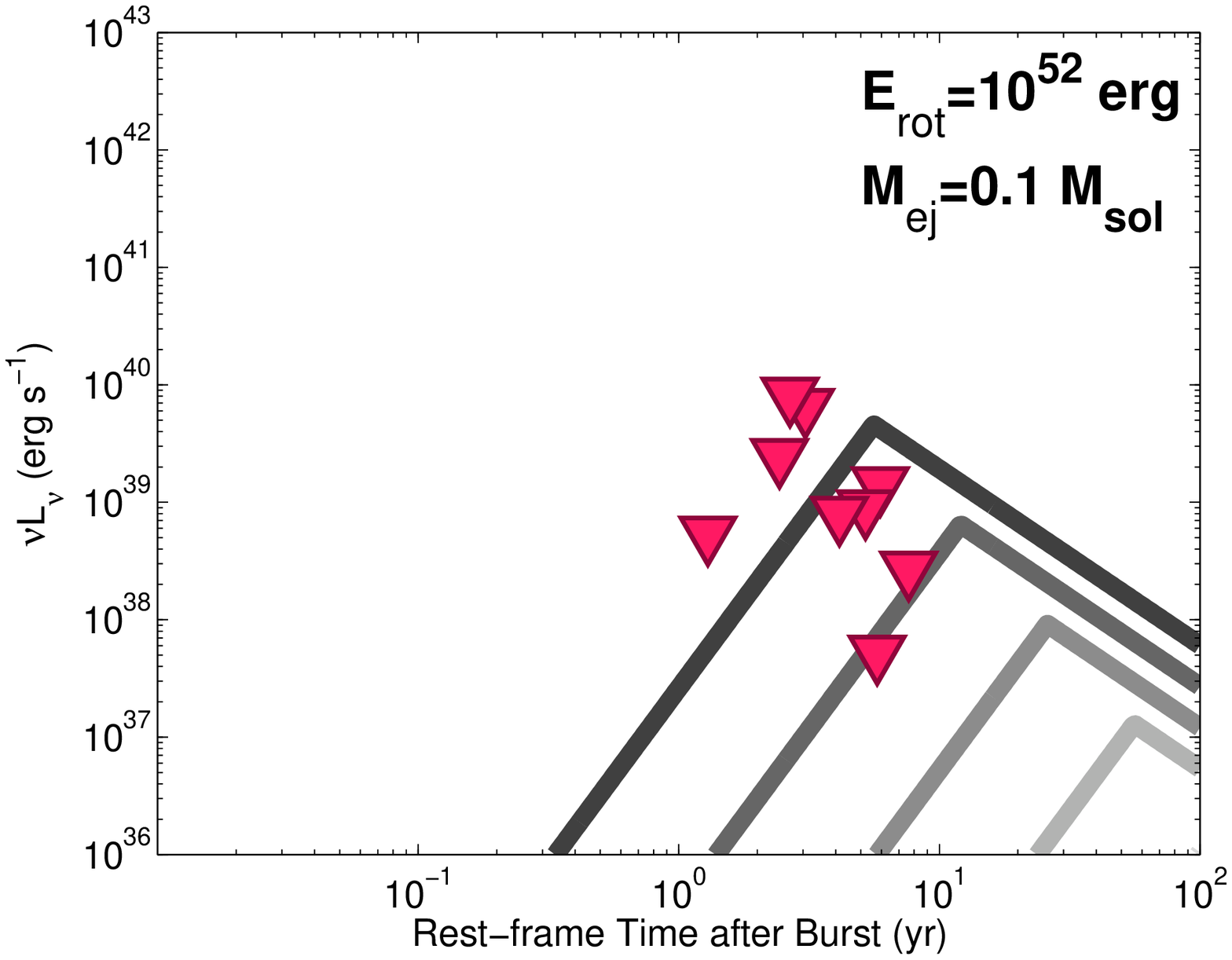} \\
\end{minipage}
\vspace{-0.15in}
\caption{$6.0$~GHz luminosity upper limits placed by VLA observations at the positions of nine short GRBs (red triangles), where the limits correspond to $3\sigma$ confidence. Also shown are light curve models at $6.0$~GHz for varying values of circumburst density ($10^{-4}-1$~cm$^{-3}$; grey curves), magnetar rotational energy ($10^{52}$~erg and $10^{53}$~erg), and ejecta masses ($0.03\,~M_{\odot}$ and $0.1\,~M_{\odot}$). These models assume $p=2.4$ and $\epsilon_e=\epsilon_B=0.1$. A comparison of the VLA observations to the brightest models ($10^{53}$~erg, $M_{\rm ej}=0.03\,~M_{\odot}$) requires that the circumburst densities be $\lesssim (0.09-5) \times 10^{-3}$~cm$^{-3}$ to accommodate a $10^{53}$~erg magnetar. Similarly, a comparison to the faintest models ($10^{52}$~erg, $M_{\rm ej}=0.1\,~M_{\odot}$) contrains the circumburst densities to $\lesssim 0.1-5$~cm$^{-3}$.
\label{fig:lc}}
\end{figure*}

\bigskip

\section{Analysis \& Results}
\label{sec:ar}

\subsection{Light curves}
\label{sec:lc}

Using Equations~\ref{eqn:tdec}$-$\ref{eqn:fobs}, we calculate a suite of model light curves for a range of magnetar energies, ejecta masses, and circumburst densities. For the magnetar energy, we consider the maximum available rotational energy of $10^{53}$~erg, corresponding to a stable or nearly stable magnetar with a relatively low mass, i.e, $M_{\rm ns} \lesssim 2.2~M_{\odot}$ \citep{mmk+15}. We also consider a more conservative energy of $10^{52}$~erg, which corresponds to the rotational energy that is removed prior to black hole formation for a supramassive NS with a mass $\approx 10\%$ higher than the maximum mass of a non-rotating NS, i.e., $M_{\rm ns} \approx 2.5~M_{\odot}$ \citep{mmk+15}.  

Motivated by numerical simulations showing that long-lived merger remnants eject a high percentage, $\gtrsim 30\%$, of the remnant accretion disk mass \citep{mf14}, we consider fiducial values for the ejecta mass of $0.03\,~M_{\odot}$ and $0.1\,~M_{\odot}$. Short GRBs explode in low-density environments, with a median circumburst density of $\approx 4 \times 10^{-3}$~cm$^{-3}$ \citep{fbm+15}; thus we consider a range of circumburst densities over $10^{-4}-1$~cm$^{-3}$. We fix the value of the power-law index $p$ to the median of the short GRB population, $p=2.4$ \citep{fbm+15}, $\epsilon_e=\epsilon_B=0.1$, and $\nu_{\rm obs}=6.0$~GHz. The resulting light curves are shown in Figure~\ref{fig:lc}. To compare the model light curves to the observations, we use the luminosity distance and redshift of each burst to convert the radio flux densities to luminosity and the observed time to rest-frame time after the burst ($\delta t_{\rm rest}$). The data are shown in Figure~\ref{fig:lc} and listed in Table~\ref{tab:obs}.

Our observations span $\delta t_{\rm rest} \approx 1.3-7.6$~yr and the luminosity limits range between $\nu L_{\nu} \lesssim (0.05-8.3) \times 10^{39}$~erg~s$^{-1}$ (Table~\ref{tab:obs} and Figure~\ref{fig:lc}). The brightest models are for $E_{\rm rot}=10^{53}$~erg and $M_{\rm ej}=0.03\,~M_{\odot}$, and the timing of the observations is well-matched to the deceleration timescale for the low-density models, $t_{\rm dec} \approx 3-14$~yr for $n=10^{-4}-10^{-2}$~cm$^{-3}$ (Equation~\ref{eqn:tdec}). Consequently, the observations provide the most stringent limits on the circumburst densities, and require that $n\lesssim (0.091-4.8) \times 10^{-3}$~cm$^{-3}$ for such a magnetar to be present. However, for a higher ejecta mass of $0.1\,~M_{\odot}$, the deceleration timescale is prolonged by a factor of $\approx 3$ (Equation~\ref{eqn:tdec}) and the observations are primarily sensitive to the rising portion of the lightcurves for the lower-density models. We therefore find less stringent circumburst density constraints of $n \lesssim (1.0-55) \times 10^{-3}$~cm$^{-3}$ (Figure~\ref{fig:lc}).

Lowering the energy by a factor of 10 to $E_{\rm rot}=10^{52}$~erg lowers the peak flux by two orders of magnitude, and prolongs the deceleration time by a factor of $\approx 3$ for a given ejecta mass. For $E_{\rm rot}=10^{52}$~erg and $M_{\rm ej}=0.03\,~M_{\odot}$, the time of observations are similar to the deceleration timescale for the higher-density models, and the density constraints are $n \lesssim (0.96-47) \times 10^{-2}$~cm$^{-3}$. The faintest models are for $10^{52}$~erg and $M_{\rm ej}=0.1\,~M_{\odot}$; the circumburst density requirements are $n\lesssim 0.11-5.9$~cm$^{-3}$.

Overall, considering a magnetar which injects $10^{53}$~erg of rotational energy, our observations uniformly rule out models for $n\gtrsim 4.8 \times 10^{-3}$~cm$^{-3}$ ($n \gtrsim 0.06$~cm$^{-3}$) for an ejecta mass of $0.03\,~M_{\odot}$ ($0.1\,~M_{\odot}$). Similarly, in the more conservative case of $10^{52}$~erg of energy output, the observations rule out models for $n \gtrsim 0.47$~cm$^{-3}$ ($\gtrsim 5.9$~cm$^{-3}$) for an ejecta mass of $0.03\,~M_{\odot}$ ($0.1\,~M_{\odot}$).

\tabletypesize{\small}
\linespread{1.3}
\begin{deluxetable*}{lccc}
\tablecolumns{4}
\tablewidth{0pc}
\tablecaption{Constraints on Magnetar Properties
\label{tab:param}}
\tablehead {
\colhead {GRB}                &
\colhead {Circumburst Density$^{a}$}                &
\colhead {$E_{\rm max}^{b}$ ($M_{\rm ej}=0.03~M_{\odot}$)}            &
\colhead {$E_{\rm max}^{b}$ ($M_{\rm ej}=0.1~M_{\odot}$)}            \\
\colhead {}                    &
\colhead {(cm$^{-3}$)}                    &
\colhead {(erg)}         &
\colhead {(erg)}                                 
}
\startdata
GRB\,050724A & $0.89^{+0.58}_{-0.49}$ & $2.2^{+0.36}_{-0.31} \times 10^{51}$ & $4.7^{+1.8}_{-1.3} \times 10^{51}$  \\
GRB\,051221A & $0.03^{+0.006}_{-0.005}$ & $1.3^{+0.19}_{-0.17} \times 10^{52}$ & $4.5^{+0.43}_{-0.39} \times 10^{52}$ \\
GRB\,070724A & $1.9^{+12}_{-1.6} \times 10^{-5}$ & $4.8^{+7.6}_{-3.0} \times 10^{53}$ & $1.6^{+2.5}_{-0.98} \times 10^{54}$  \\
GRB\,080905A & $1.3^{+33}_{-1.2} \times 10^{-4}$ & $7.0^{+23}_{-5.4} \times 10^{52}$ & $2.3^{+7.5}_{-1.8} \times 10^{53}$ \\
GRB\,090510 & $1.2^{+5.5}_{-1.0} \times 10^{-5}$ & $1.7^{+2.3}_{-9.6} \times 10^{54}$ & $5.5^{+7.7}_{-3.2} \times 10^{54}$ \\
GRB\,090515 & \nod & \nod & \nod \\
GRB\,100117A & $0.04^{+0.03}_{-0.01}$ & $3.3^{+0.77}_{-6.2} \times 10^{52}$ & $1.1^{+0.26}_{-0.21} \times 10^{53}$ \\
GRB\,101219A & $4.6^{+59}_{-4.3} \times 10^{-5}$ & $8.0^{+21.8}_{-5.8} \times 10^{53}$ & $2.6^{+7.3}_{-1.9} \times 10^{54}$ \\
GRB\,130603B & $0.09^{+0.04}_{-0.03}$ & $2.1^{+0.44}_{-0.36} \times 10^{52}$ & $6.9^{+1.5}_{-1.2} \times 10^{52}$
\enddata
\tablecomments{Reported uncertainties correspond to $1\sigma$ confidence. \\
$^{a}$ Inferred circumburst densities as determined from afterglow observations \citep{fbm+15}. \\
$^{b}$ Maximum energy of a magnetar allowed by the observations assuming $\epsilon_B=0.1$.}
\end{deluxetable*}

\linespread{1.0}

\begin{figure*}
\begin{minipage}[c]{\textwidth}
\tabcolsep0.0in
\includegraphics*[width=0.33\textwidth,clip=]{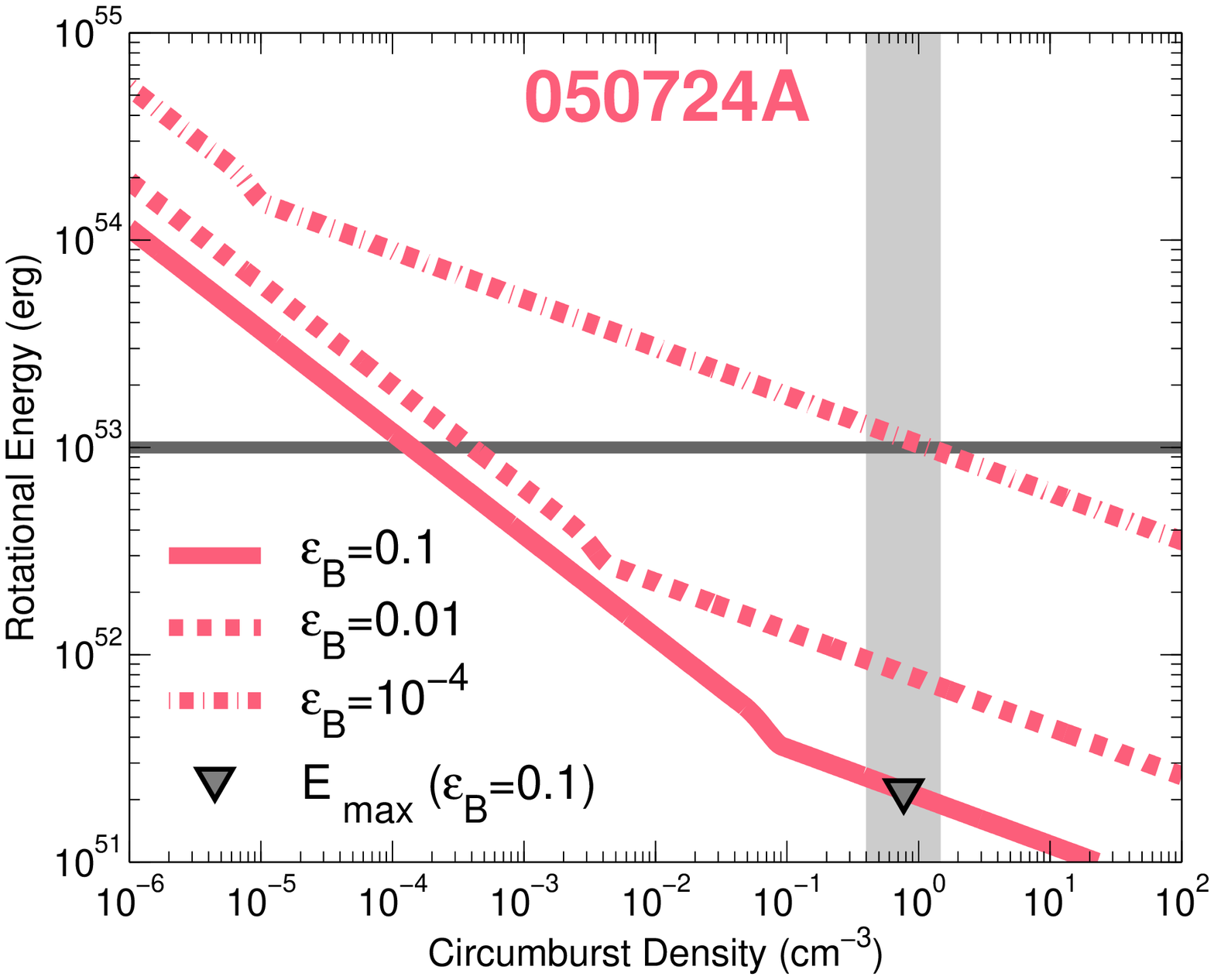}
\includegraphics*[width=0.33\textwidth,clip=]{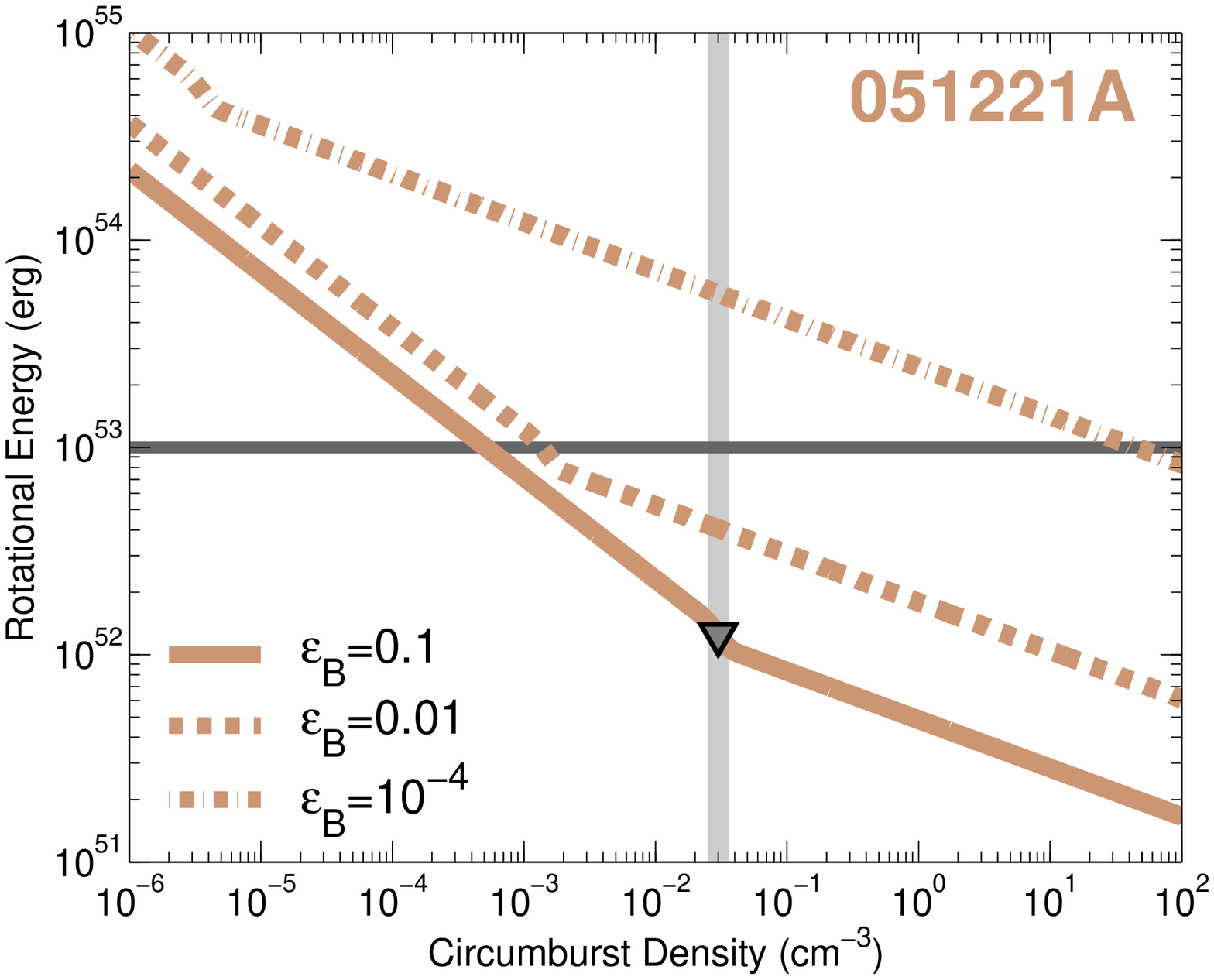}
\includegraphics*[width=0.33\textwidth,clip=]{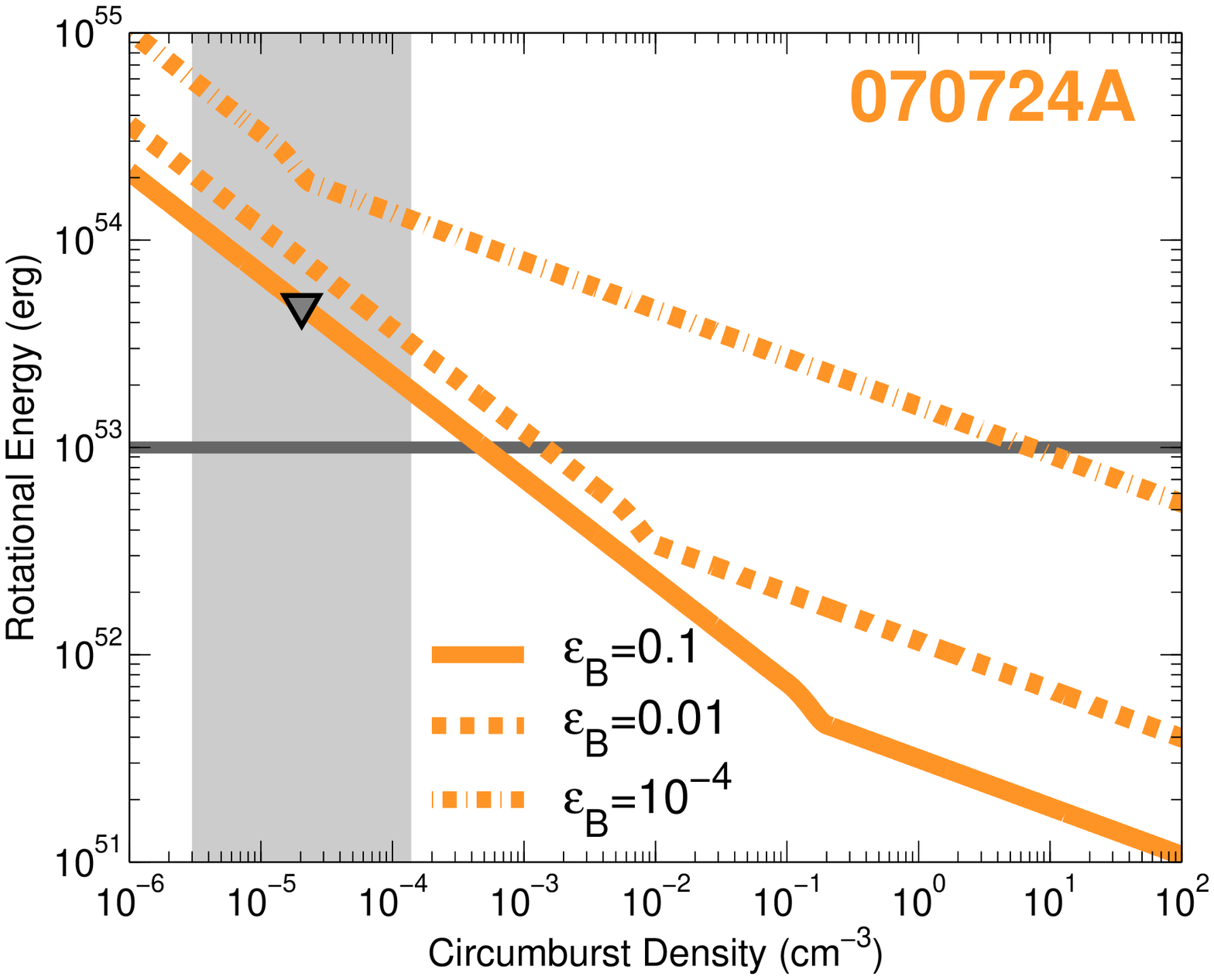} \\
\includegraphics*[width=0.33\textwidth,clip=]{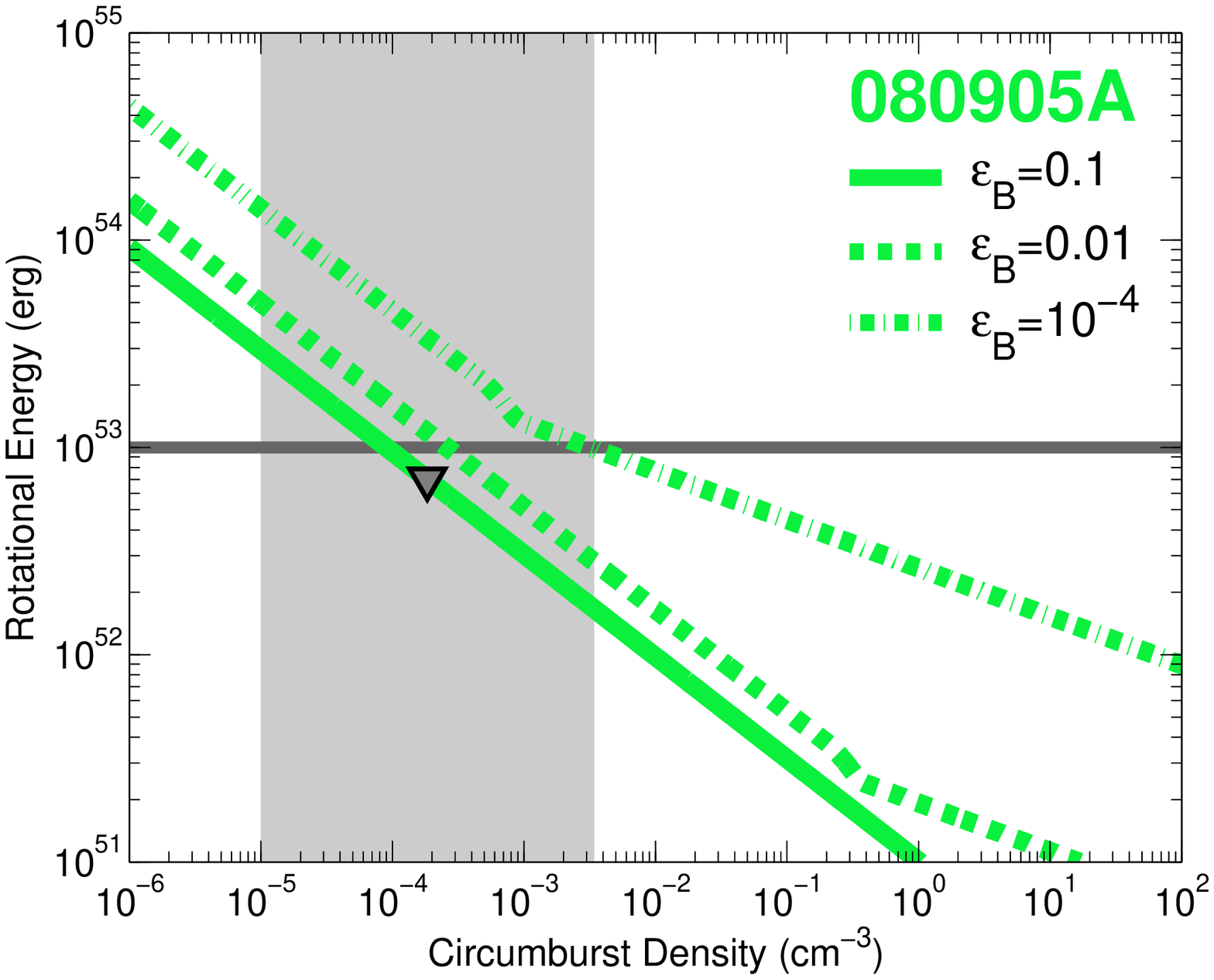}
\includegraphics*[width=0.33\textwidth,clip=]{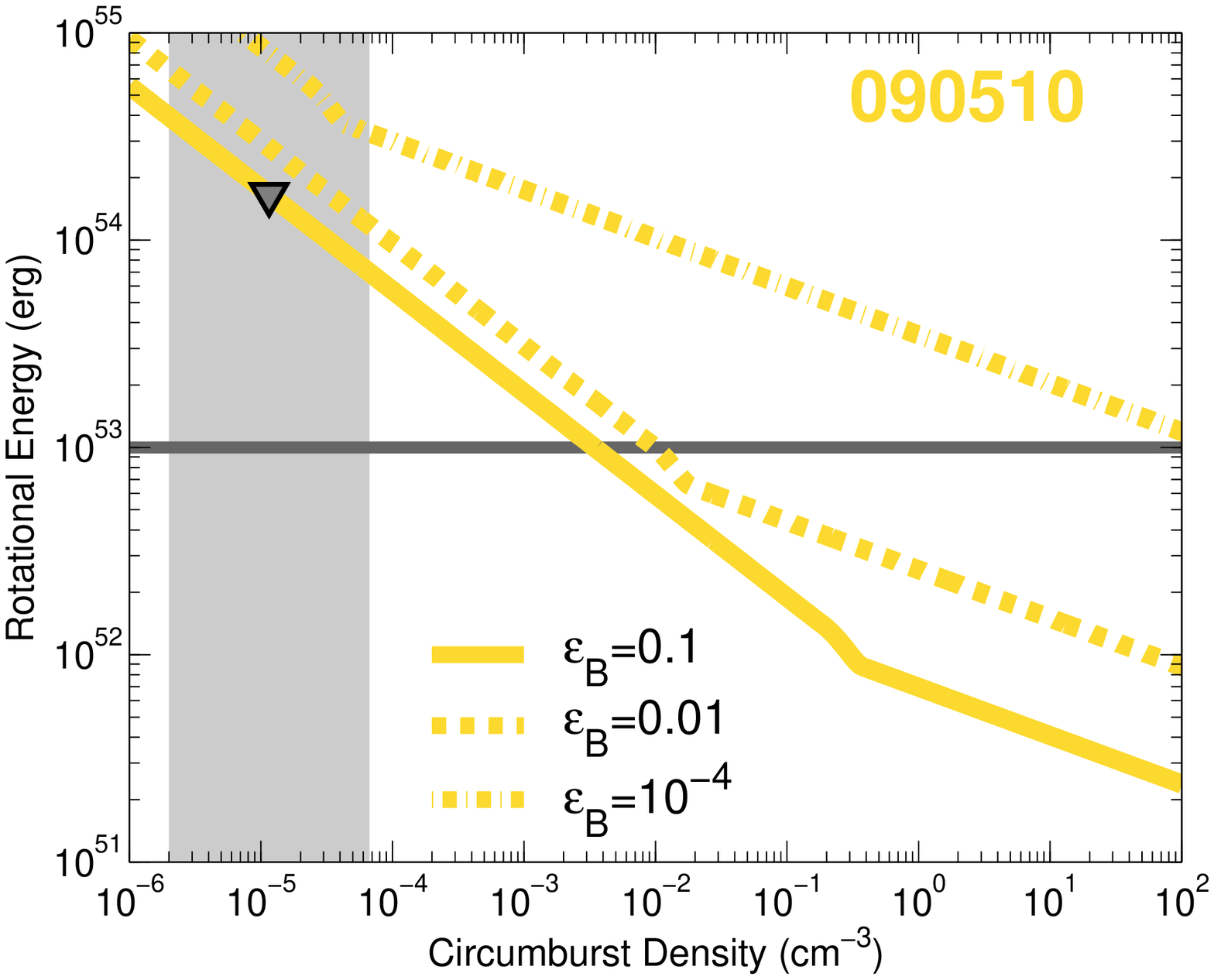}
\includegraphics*[width=0.33\textwidth,clip=]{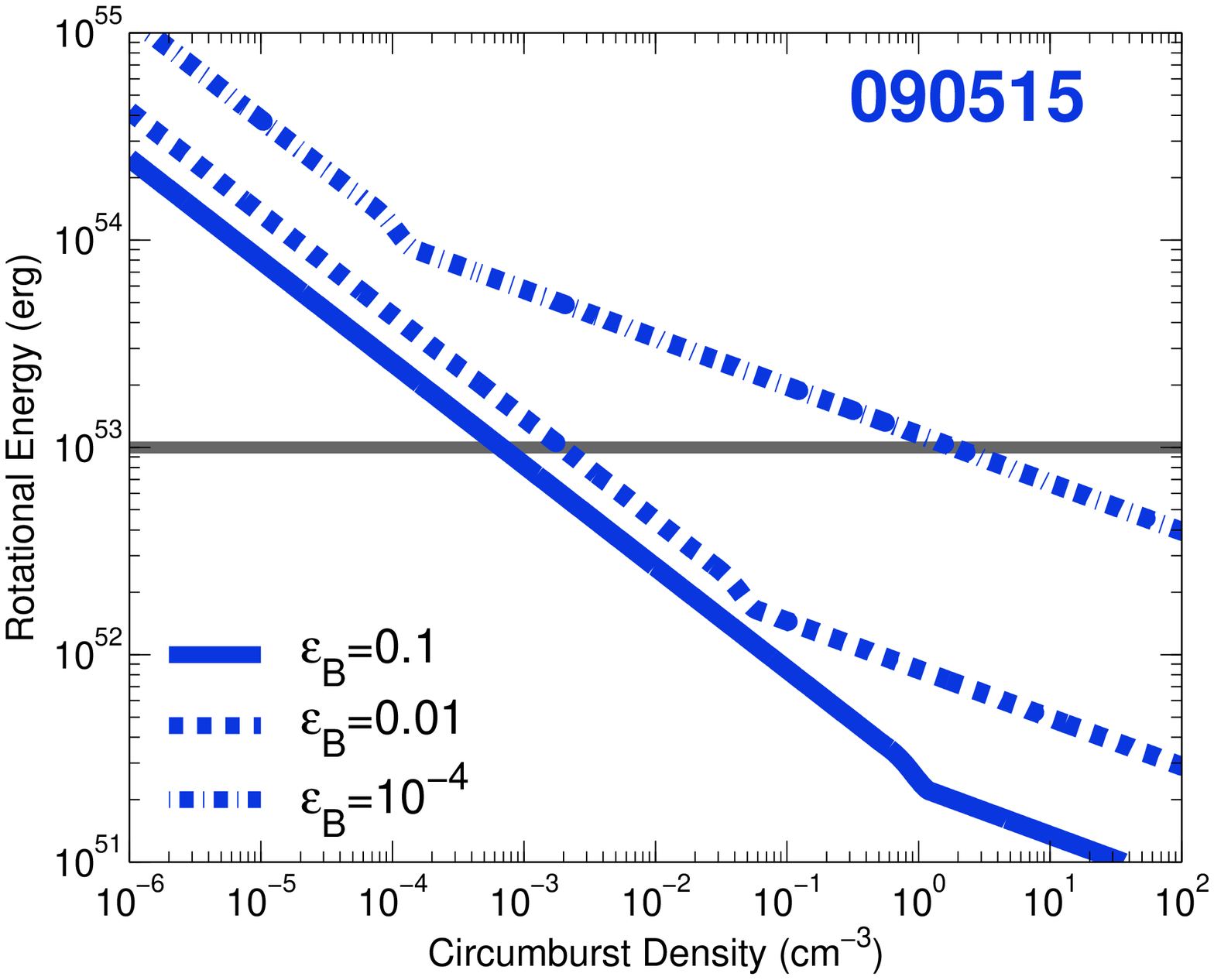} \\
\includegraphics*[width=0.33\textwidth,clip=]{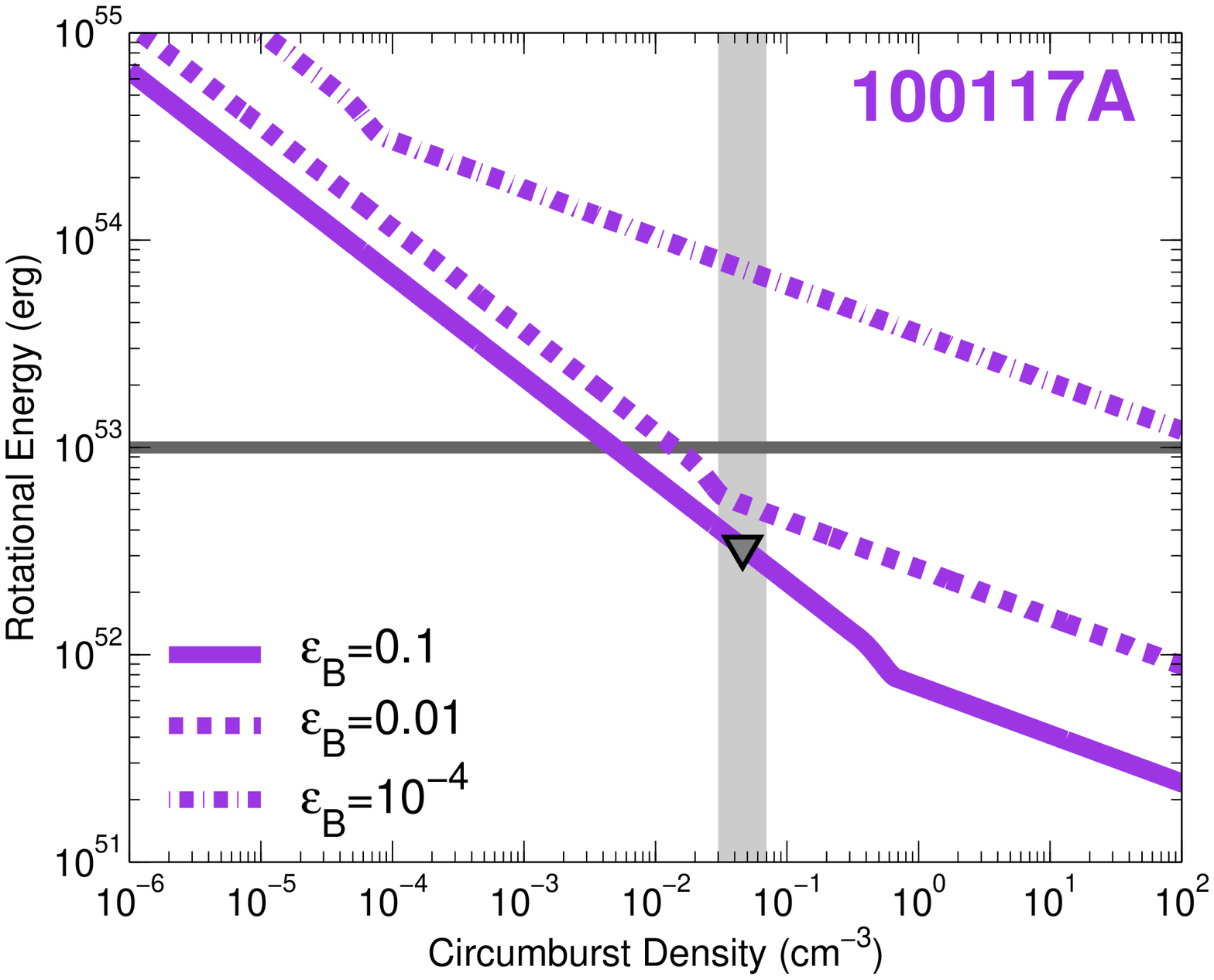}
\includegraphics*[width=0.33\textwidth,clip=]{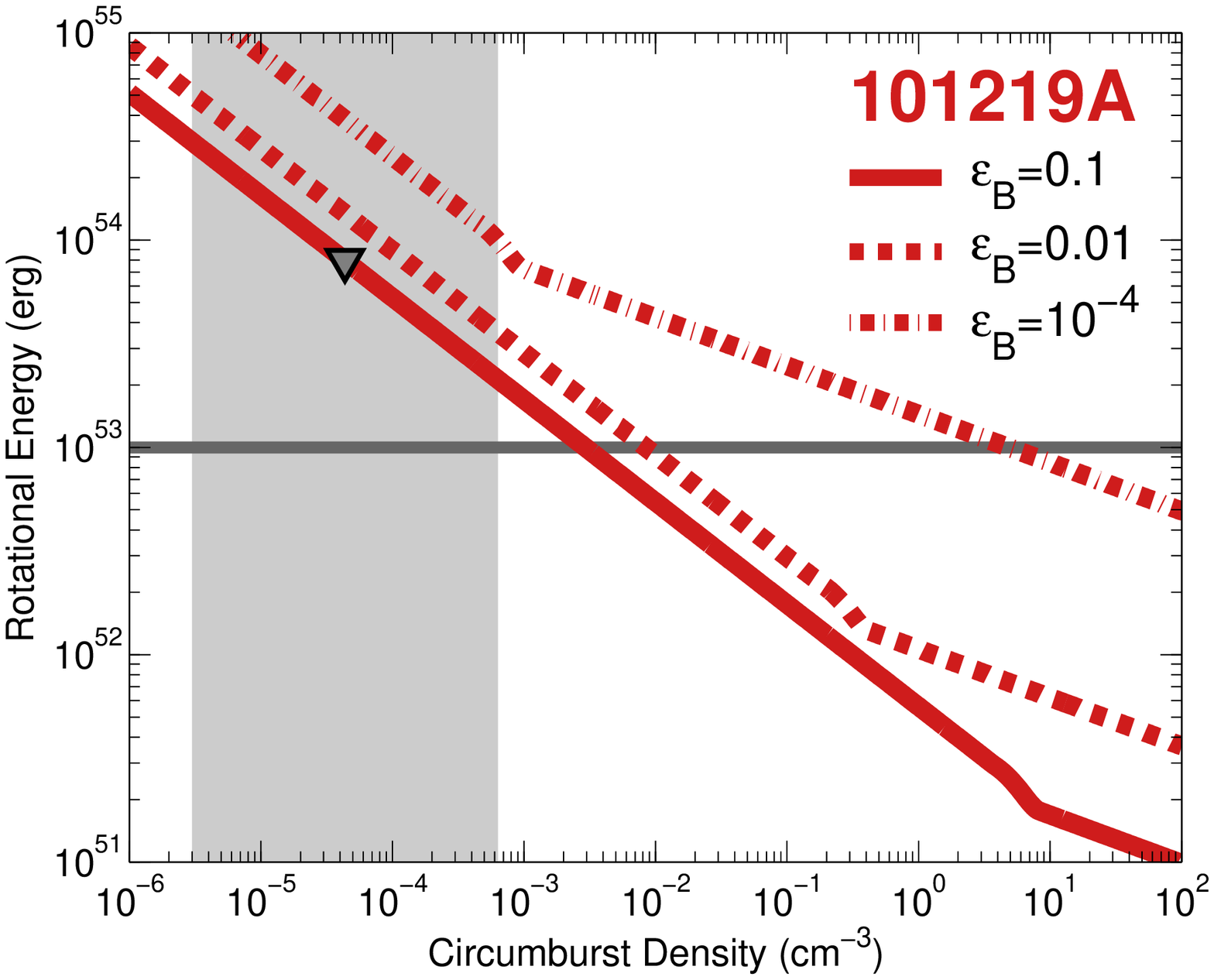}
\includegraphics*[width=0.33\textwidth,clip=]{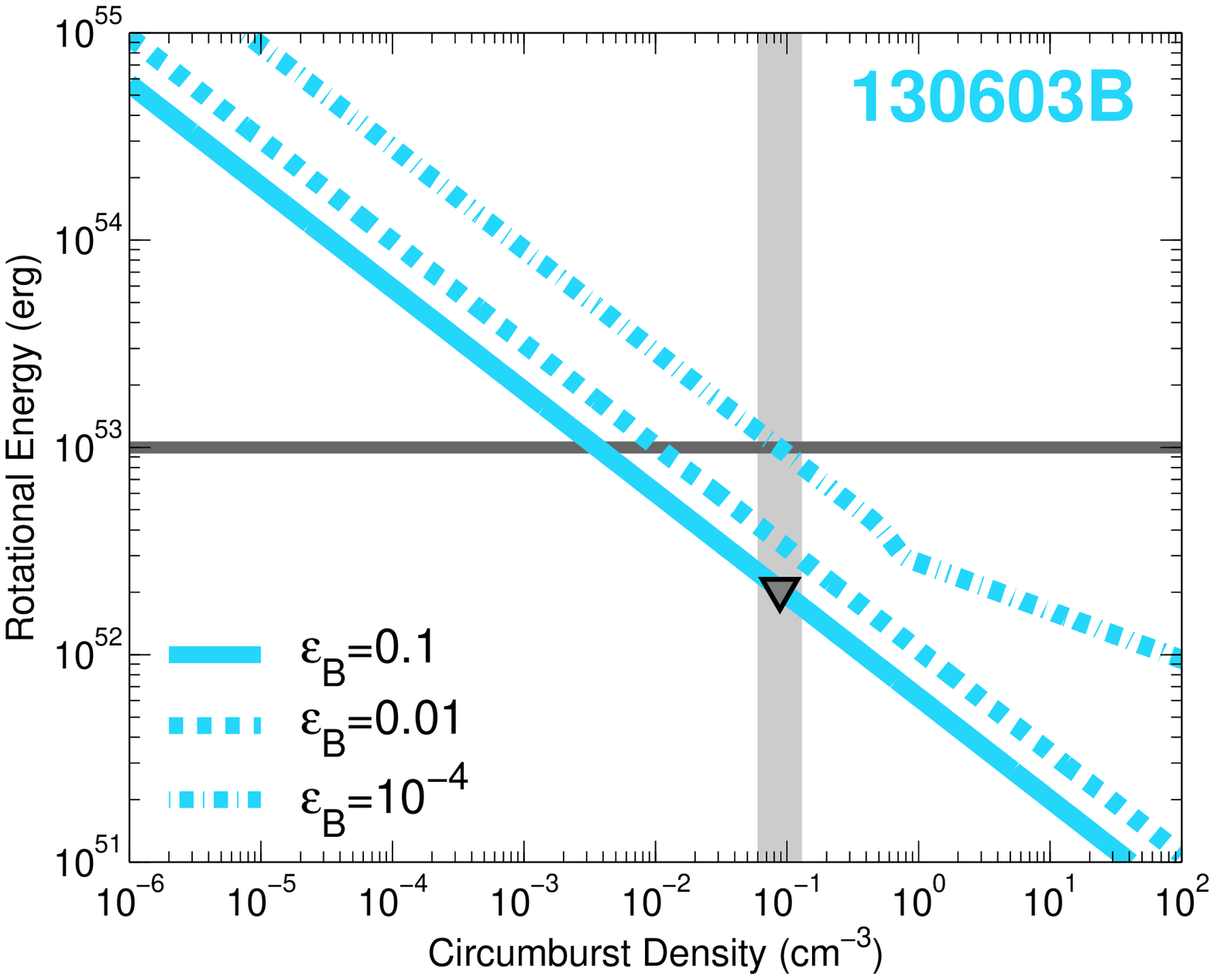}
\end{minipage}
\caption{Constraints on the rotational energy-circumburst density parameter space from VLA observations for the nine short GRBs in our sample. Constraints are shown for $M_{\rm ej}=0.03 ~M_{\odot}$ and $\epsilon_B=0.1$, $0.01$, and $10^{-4}$ (solid, dashed, and dot-dashed curves, respectively). In each panel, the curves represent upper limits on the parameter space, where the region below each curve is allowed and the region above is ruled out. Light grey regions represent $1\sigma$ ranges of allowed circumburst densities independently determined from afterglow observations for $\epsilon_B=0.1$ \citep{fbm+15}; there is not enough information to constrain the circumburst density of GRB\,090515. The average maximum value of the rotational energy constrained by the observations, $E_{\rm max}$, at $\epsilon_B=0.1$ is denoted by a grey triangle, corresponding to the values in Table~\ref{tab:param}. A grey horizonal line shows the maximum extractable rotational energy of a $\sim 2.2 ~M_{\odot}$ magnetar of $10^{53}$~erg. The observations can rule out the presence of a $\sim 2.2 ~M_{\odot}$ magnetar for GRBs\,050724A, 051221A, 080905A, 100117A and 130603B.
\label{fig:Enpanel03}}
\end{figure*}

\begin{figure*}
\begin{minipage}[c]{\textwidth}
\tabcolsep0.0in
\includegraphics*[width=0.33\textwidth,clip=]{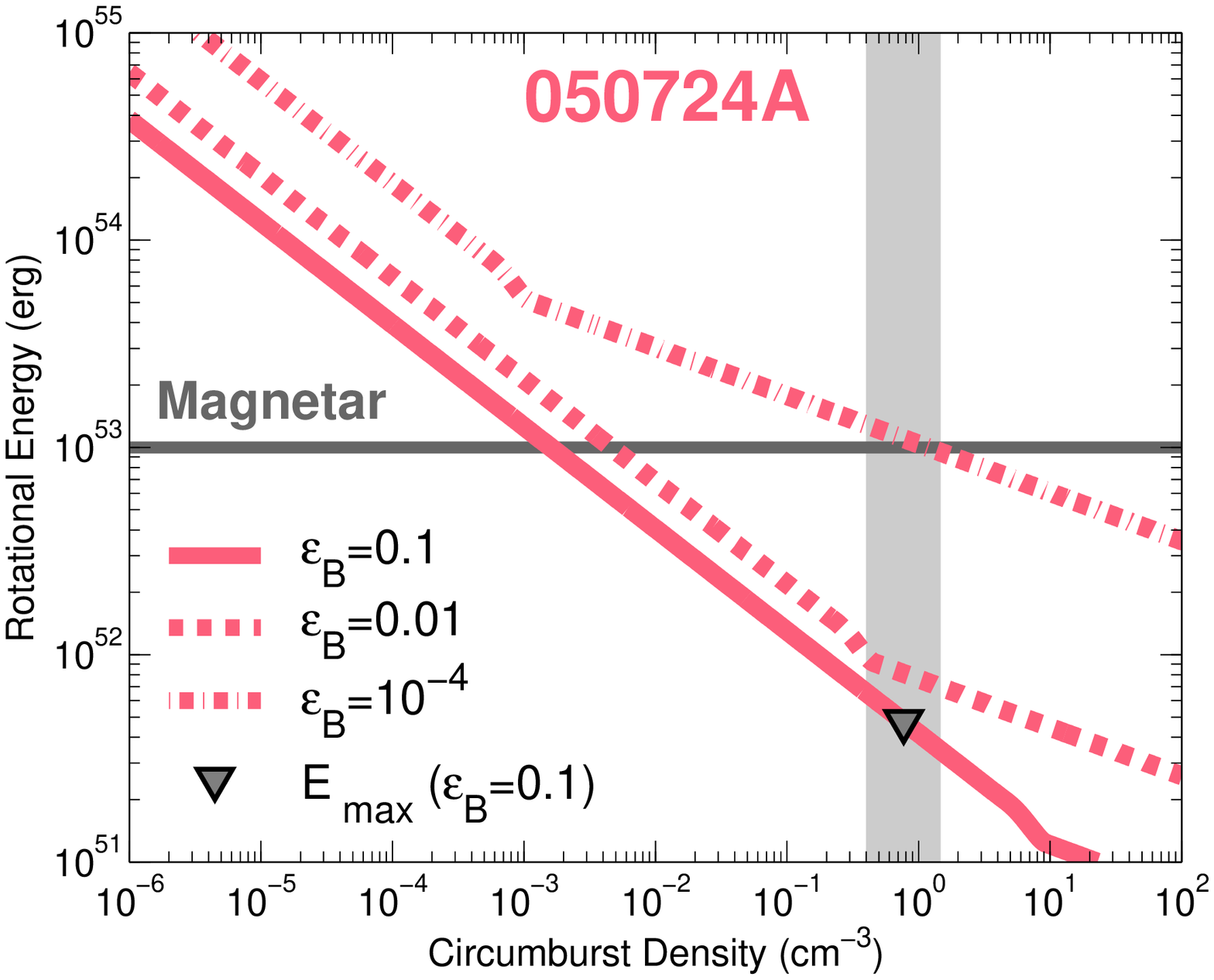}
\includegraphics*[width=0.33\textwidth,clip=]{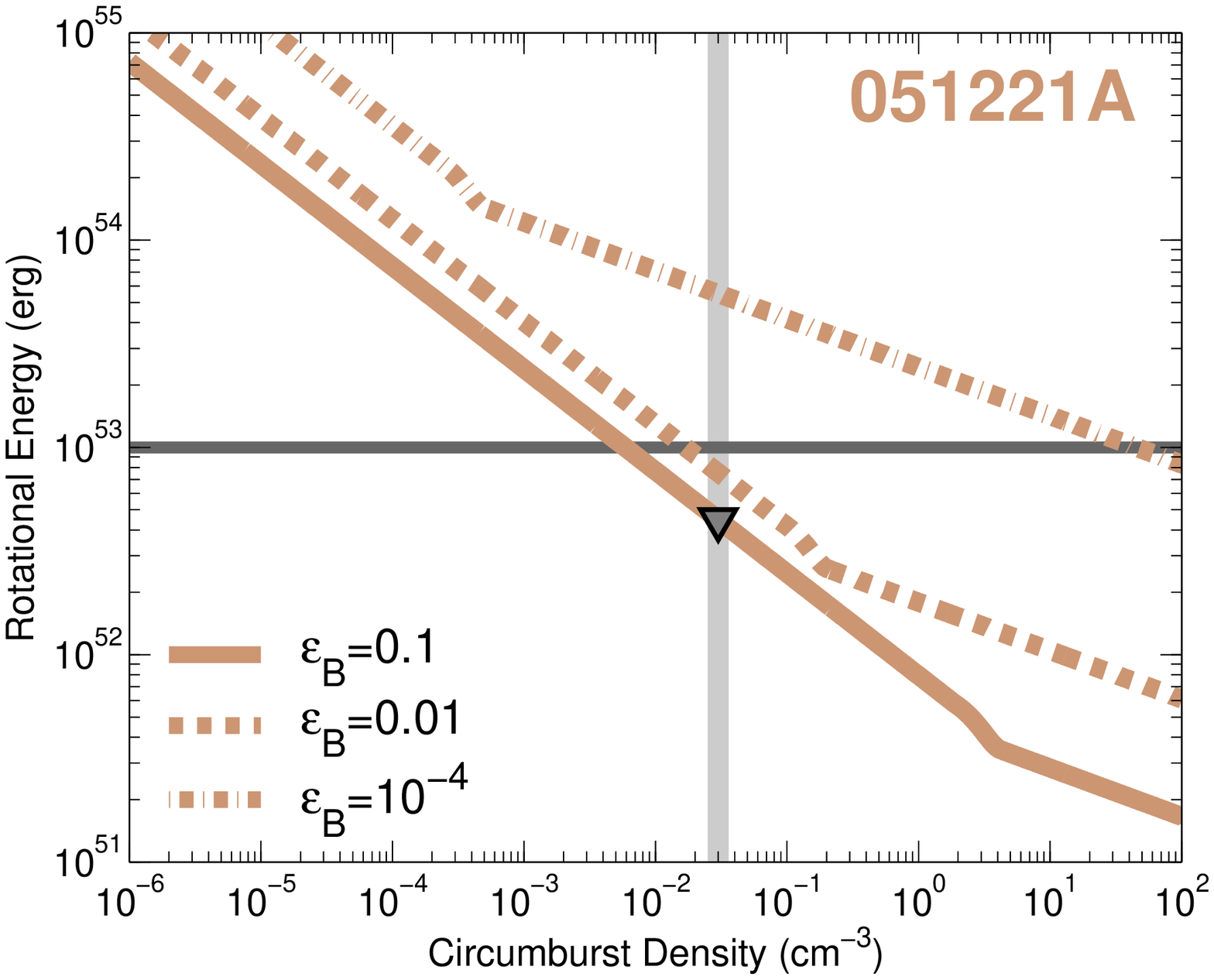}
\includegraphics*[width=0.33\textwidth,clip=]{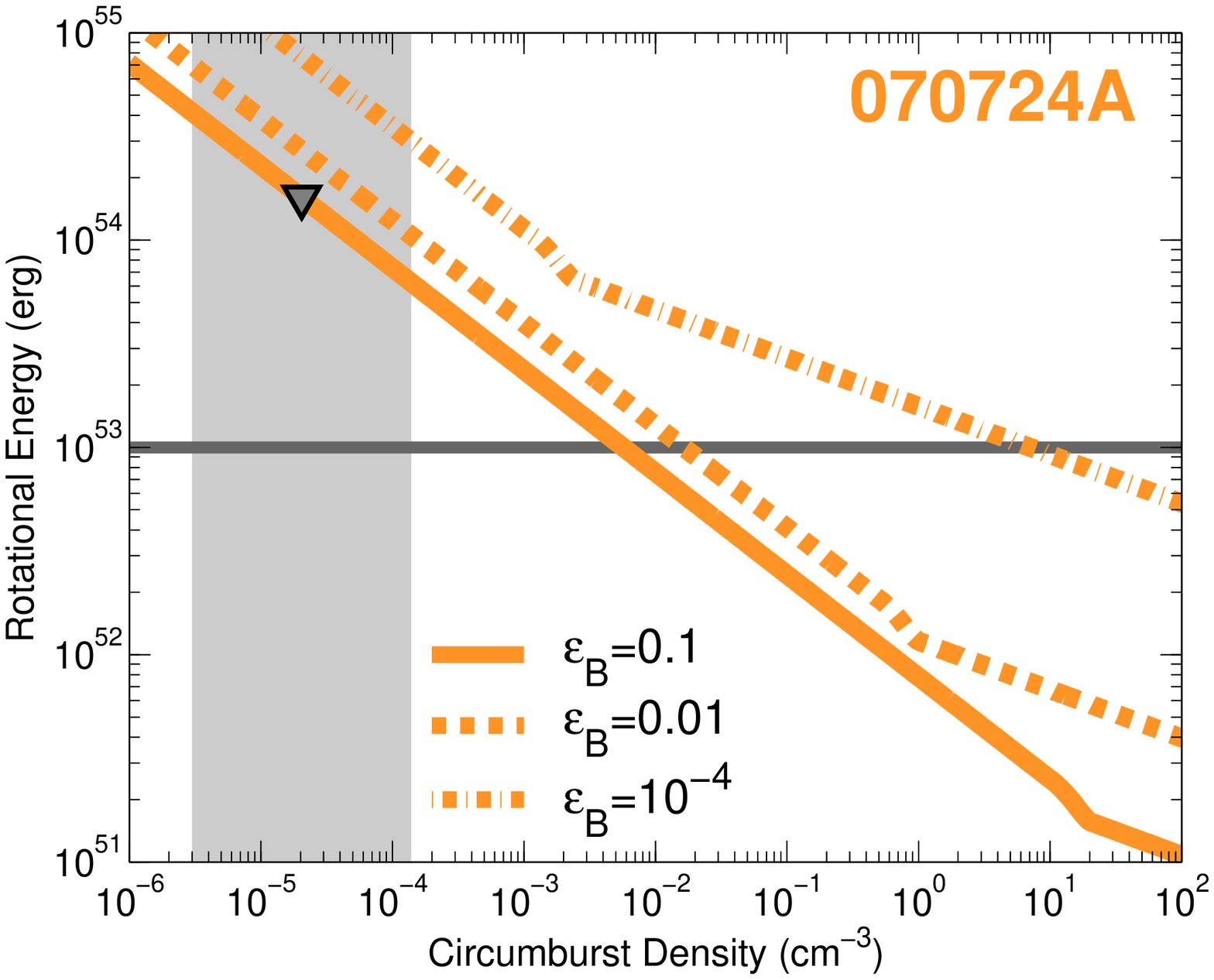} \\
\includegraphics*[width=0.33\textwidth,clip=]{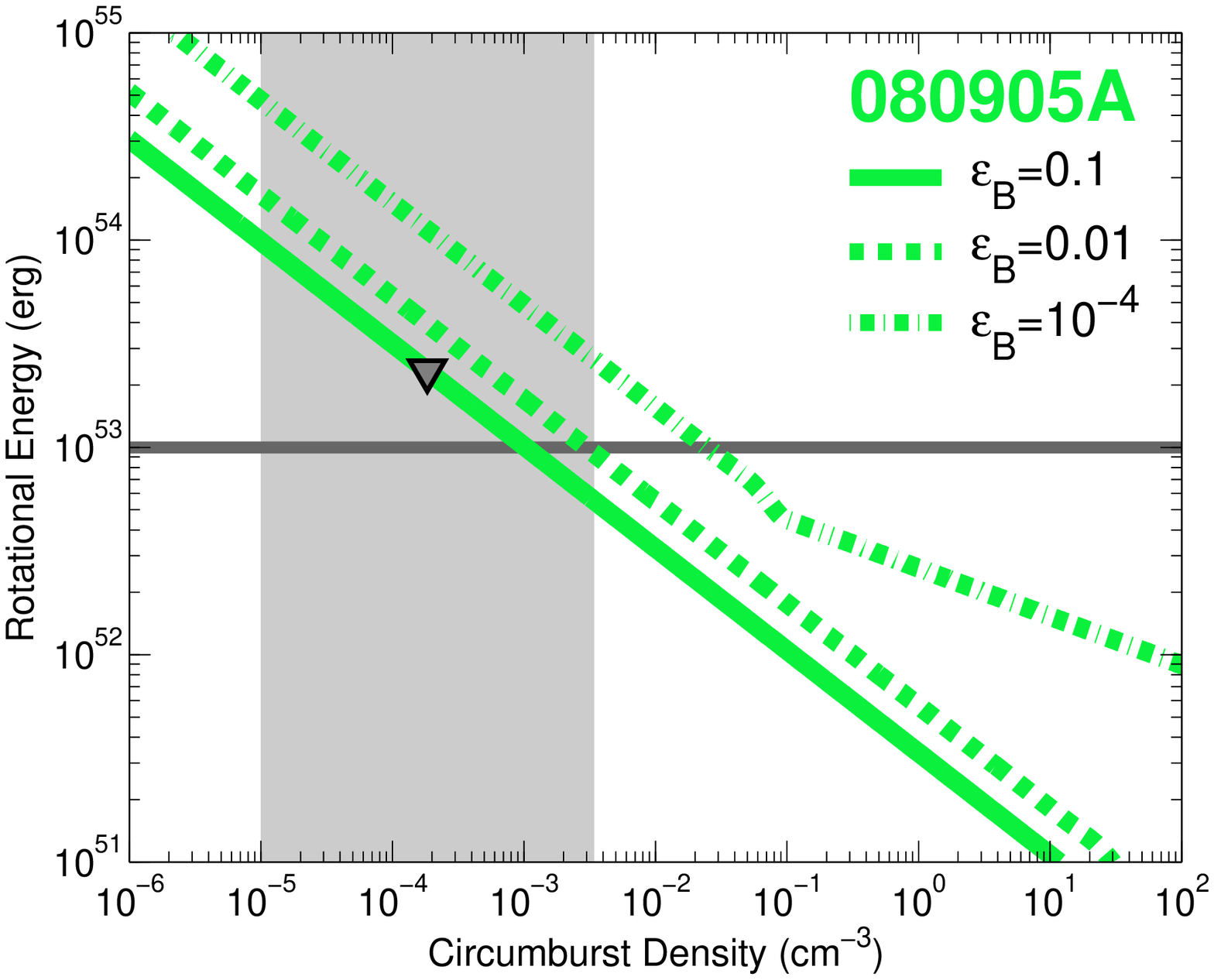}
\includegraphics*[width=0.33\textwidth,clip=]{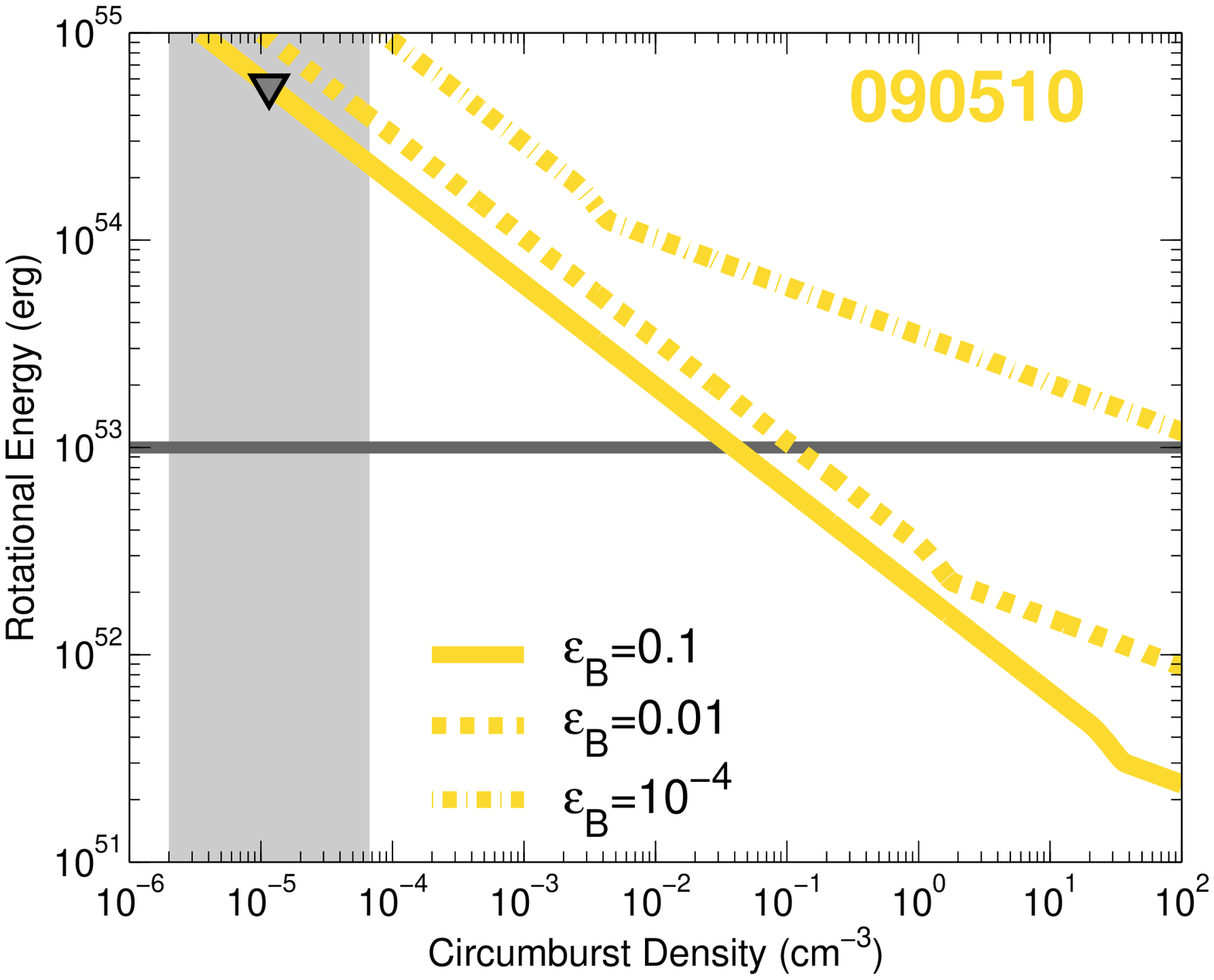}
\includegraphics*[width=0.33\textwidth,clip=]{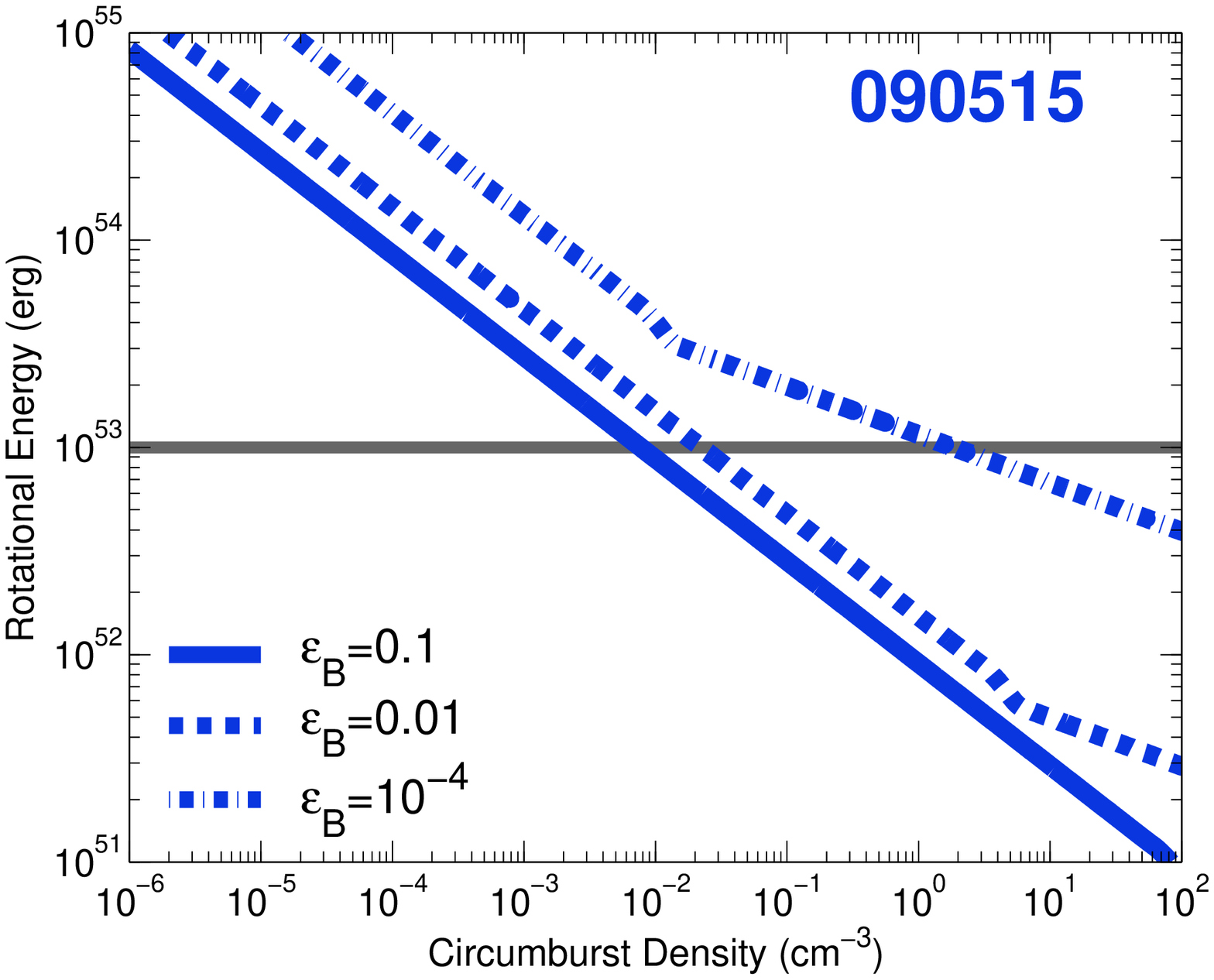} \\
\includegraphics*[width=0.33\textwidth,clip=]{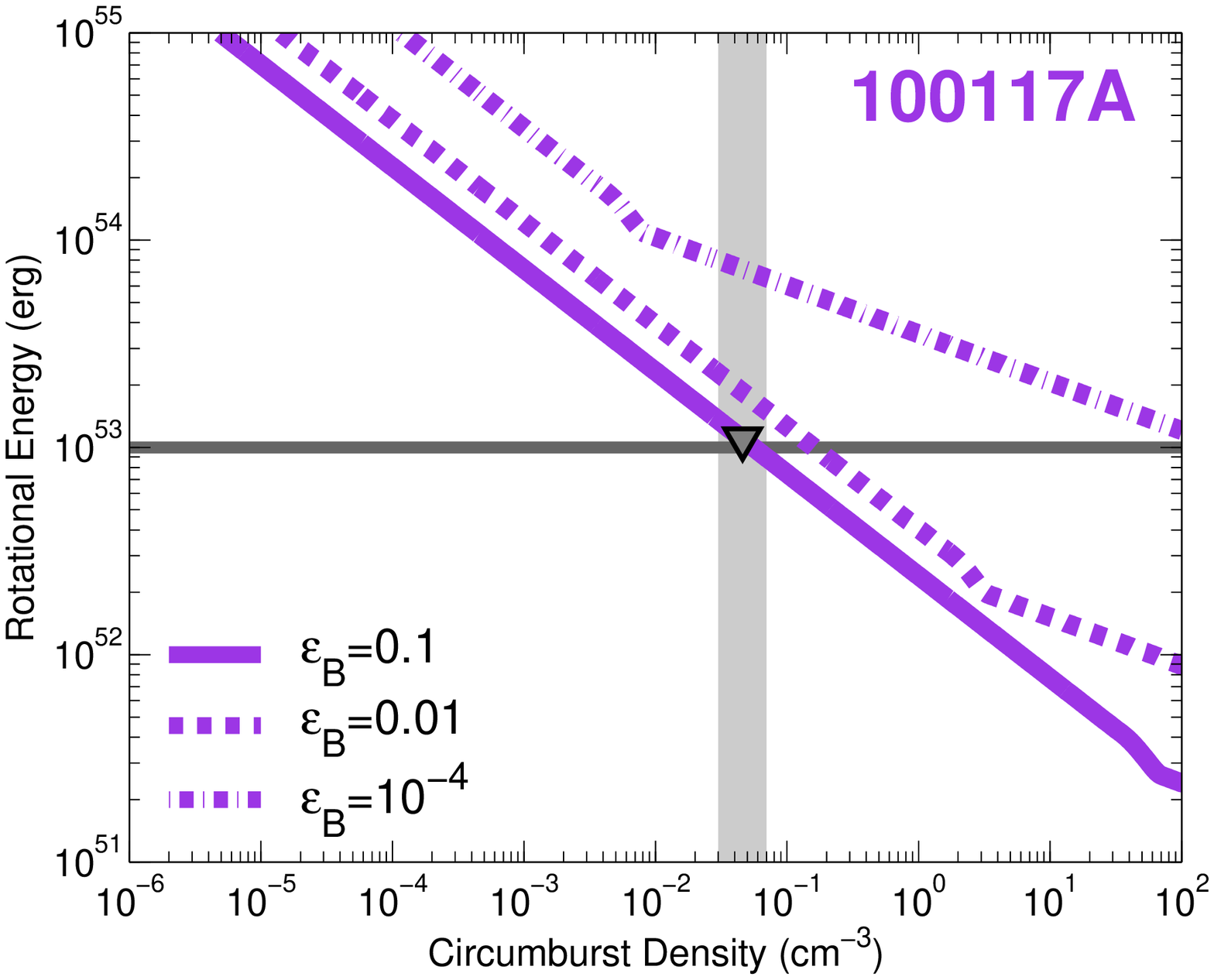}
\includegraphics*[width=0.33\textwidth,clip=]{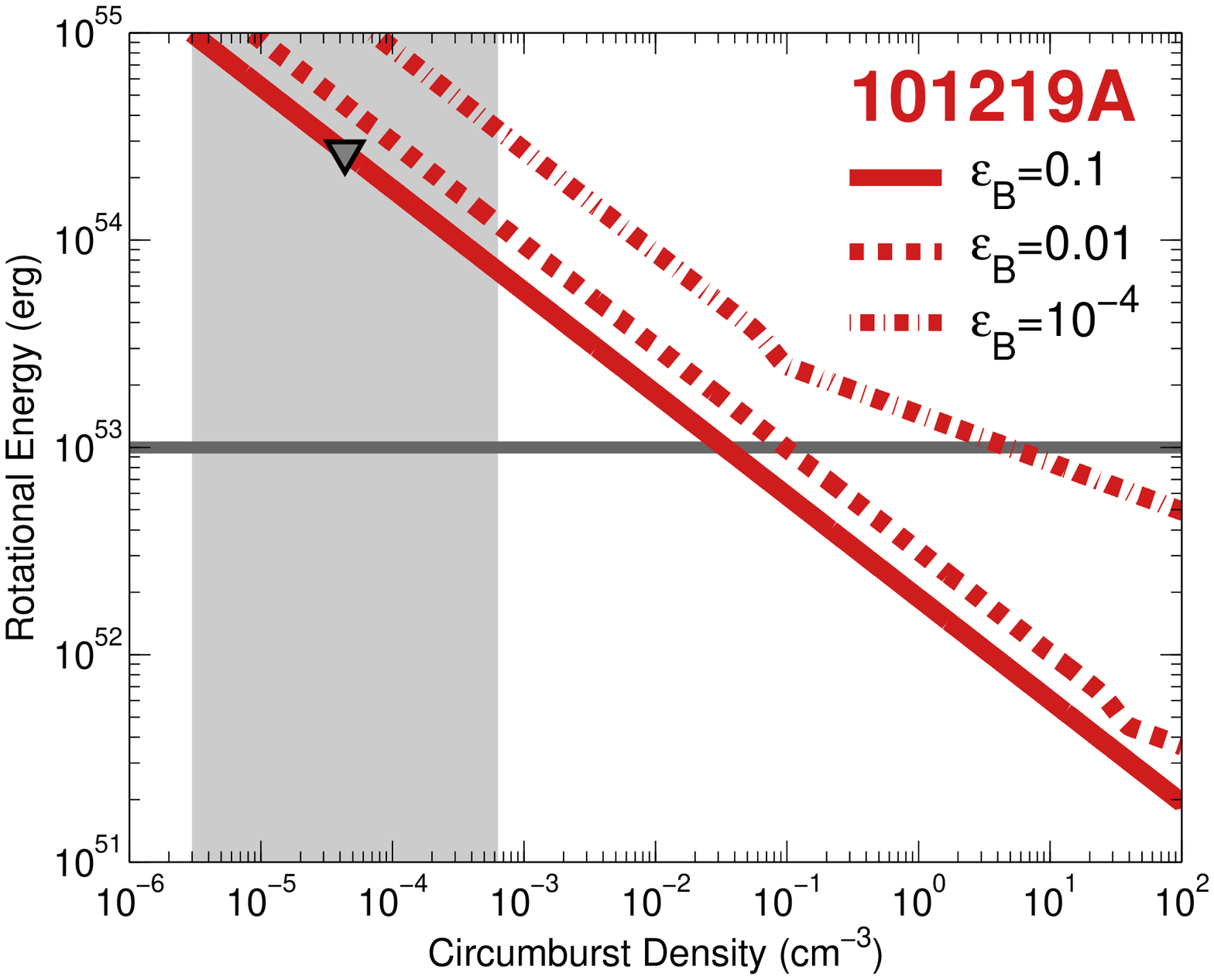}
\includegraphics*[width=0.33\textwidth,clip=]{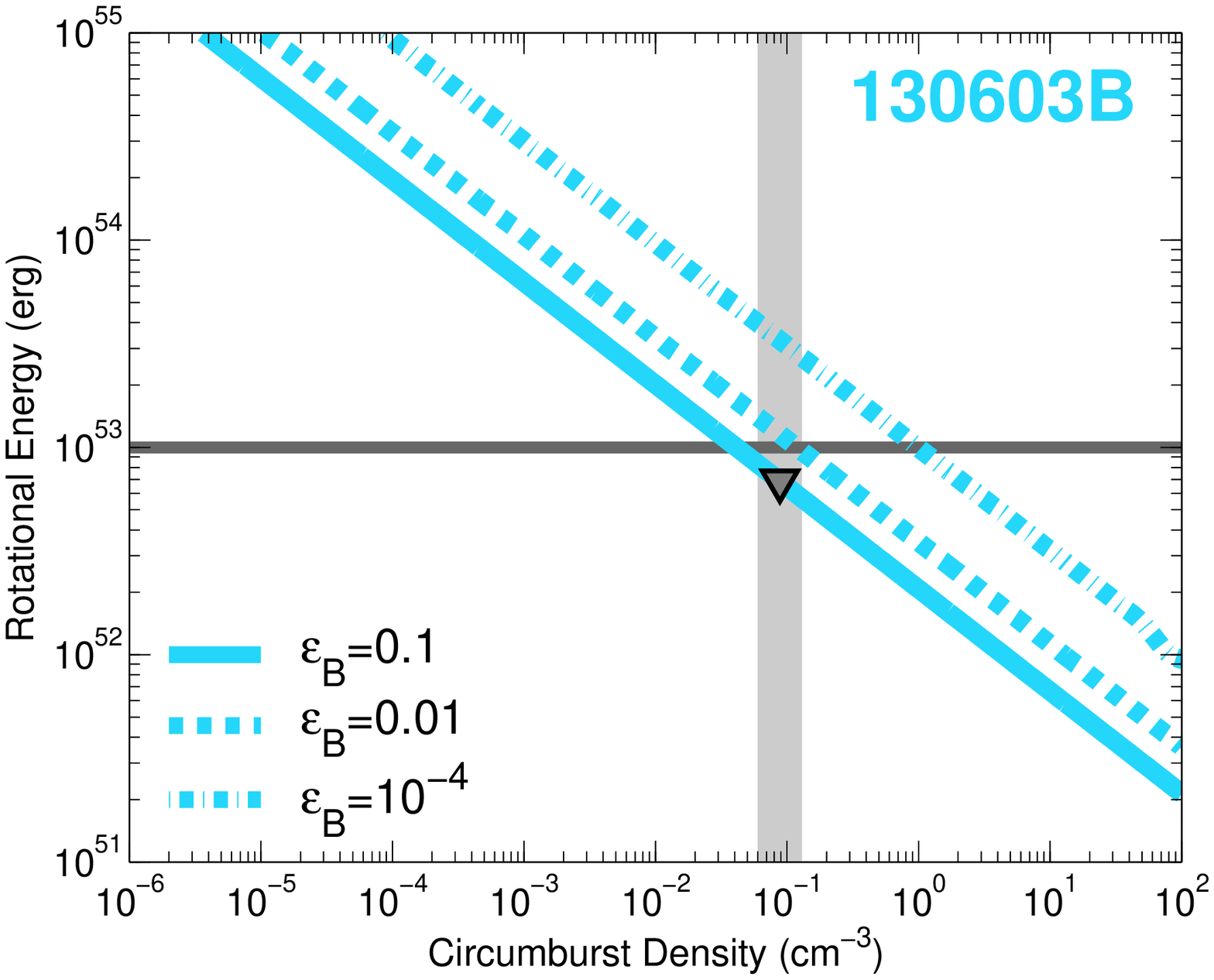}
\end{minipage}
\caption{Same as for Figure~\ref{fig:Enpanel03} but for $M_{\rm ej}=0.1 ~M_{\odot}$. The observations can rule out the presence of a $\sim 2.2 ~M_{\odot}$ magnetar for GRBs\,050724A, 051221A, 100117A and 130603B.
\label{fig:Enpanel1}}
\end{figure*}

\subsection{Maximum Magnetar Energy}

To constrain the joint rotational energy-circumburst density parameter space with the VLA observations, we use Equations~\ref{eqn:fpk}-\ref{eqn:fobs} and the data in Table~\ref{tab:obs} to obtain expressions for energy as a function of circumburst density. For each burst, we use the upper limits on the flux density to calculate an upper limit on the combination of rotational energy and circumburst density (Equations~\ref{eqn:fpk}-\ref{eqn:fobs}). We assume the same fiducial values for $p$ and $\epsilon_e$ as in Section~\ref{sec:lc}, and consider three values for $\epsilon_B$: $10^{-4}$, $0.01$ and $0.1$. The resulting upper limits on the parameter space are shown for $M_{\rm ej}=0.03\,~M_{\odot}$ and $0.1\,~M_{\odot}$ in Figure~\ref{fig:Enpanel03} and Figure~\ref{fig:Enpanel1}, respectively. The bend in each constraint represents the different behavior before and after the deceleration time.  

Additionally, broad-band modeling of the afterglow emission provides vital measurements of the circumburst density, which can be used as independent constraints on the allowed parameter space. We collect the inferred circumburst densities for eight bursts from \citet{fbm+15}; there is not enough information to constrain the circumburst density of GRB\,090515. These values are listed in Table~\ref{tab:param} and shown in Figures~\ref{fig:Enpanel03}-\ref{fig:Enpanel1}, where the ranges shown for each region denotes the $1\sigma$ uncertainty. Comparing the models over the range of densities allowed by afterglow observations gives an upper limit on the energy of a magnetar remnant, $E_{\rm max}$, listed in Table~\ref{tab:param} for both ejecta masses.

The maximum allowed energies are uniformly lower by a factor of $\approx 3$ for the smaller ejecta mass. For the four bursts with relatively well-measured densities, GRBs\,050724A, 051221A, 100117A, and 130603B, we can rule out the presence of a magnetar with $E_{\rm rot}=10^{53}$~erg for both ejecta masses. The deepest constraints are for GRB\,050724A, for which we can place limits of $E_{\rm max}\approx(2-5) \times 10^{51}$~erg, two orders of magnitude below the rotational energy of a stable $\sim 2.2\,~M_{\odot}$ magnetar (Table~\ref{tab:param}; \citealt{mmk+15}). Three of the remaining bursts, GRBs\,070724A, 090510, and 101219A, have relatively low and uncertain inferred circumburst densities of $\lesssim 10^{-3}$~cm$^{-3}$; thus the constraints on the energy are less stringent, and we cannot rule out the presence of a magnetar with $E_{\rm rot}=10^{53}$~erg. Finally, for GRB\,080905A, a $10^{53}$~erg magnetar can be ruled out for the smaller ejecta mass only (Table~\ref{tab:param} and Figures~\ref{fig:Enpanel03}-\ref{fig:Enpanel1}). If we consider a conservative case of $E_{\rm rot}=10^{52}$~erg, only the observations for GRB\,050724A can rule out the presence of such a magnetar.

\section{Discussion}
\label{sec:disc}

Prior to this effort, only two attempts have been made to search for radio emission following short GRBs on timescales of $\sim$years. \citet{mb14} used the VLA (prior to the 2010 upgrade) to observe the fields of seven short GRBs, half of which had extended emission in the X-ray band. The observations were taken at $1.425$~GHz on timescales of $\delta t_{\rm rest}\approx 0.5-2$~yr. They detected no radio emission to $3\sigma$ limits of $\approx200-500\,\mu$Jy \citep{mb14}. A second study targeted two short GRBs with claims of associated kilonovae, GRB\,060614 \citep{jlc+15,yjl+15} and GRB\,130603B \citep{bfc13,tlf+13}, and detected no radio emission to $3\sigma$ limits of $\approx150$ and $\approx60\,\mu$Jy at rest-frame times of $\approx7.9$ and $\approx1.3$~yr, respectively \citep{hhp+16}. With flux limits of $\lesssim 18-32\,\mu$Jy for nine short GRBs, our study represents the deepest and most extensive campaign for late-time radio emission following short GRBs to date.

\begin{figure*}
\begin{minipage}[c]{\textwidth}
\tabcolsep0.0in
\includegraphics*[width=0.5\textwidth,clip=]{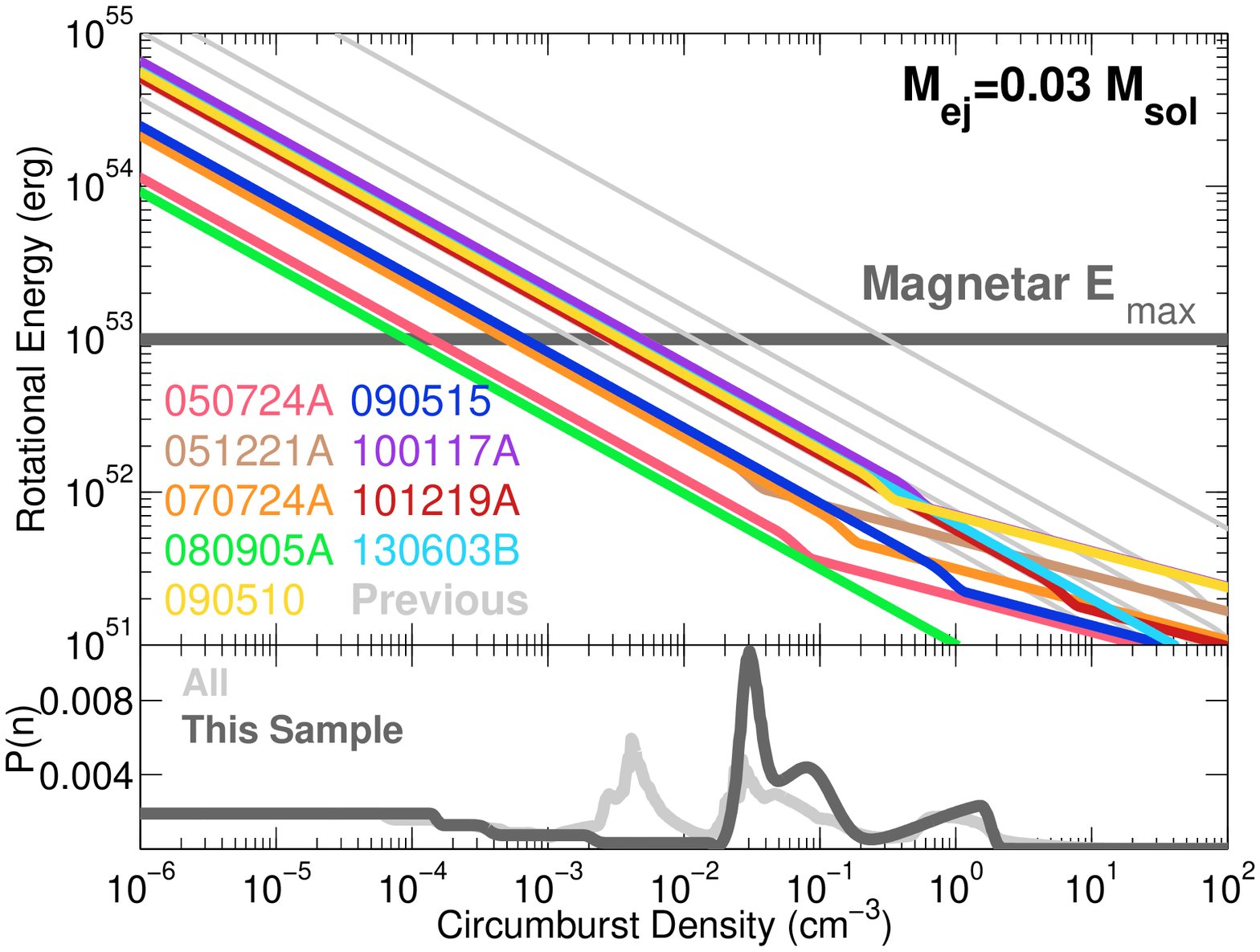}
\includegraphics*[width=0.5\textwidth,clip=]{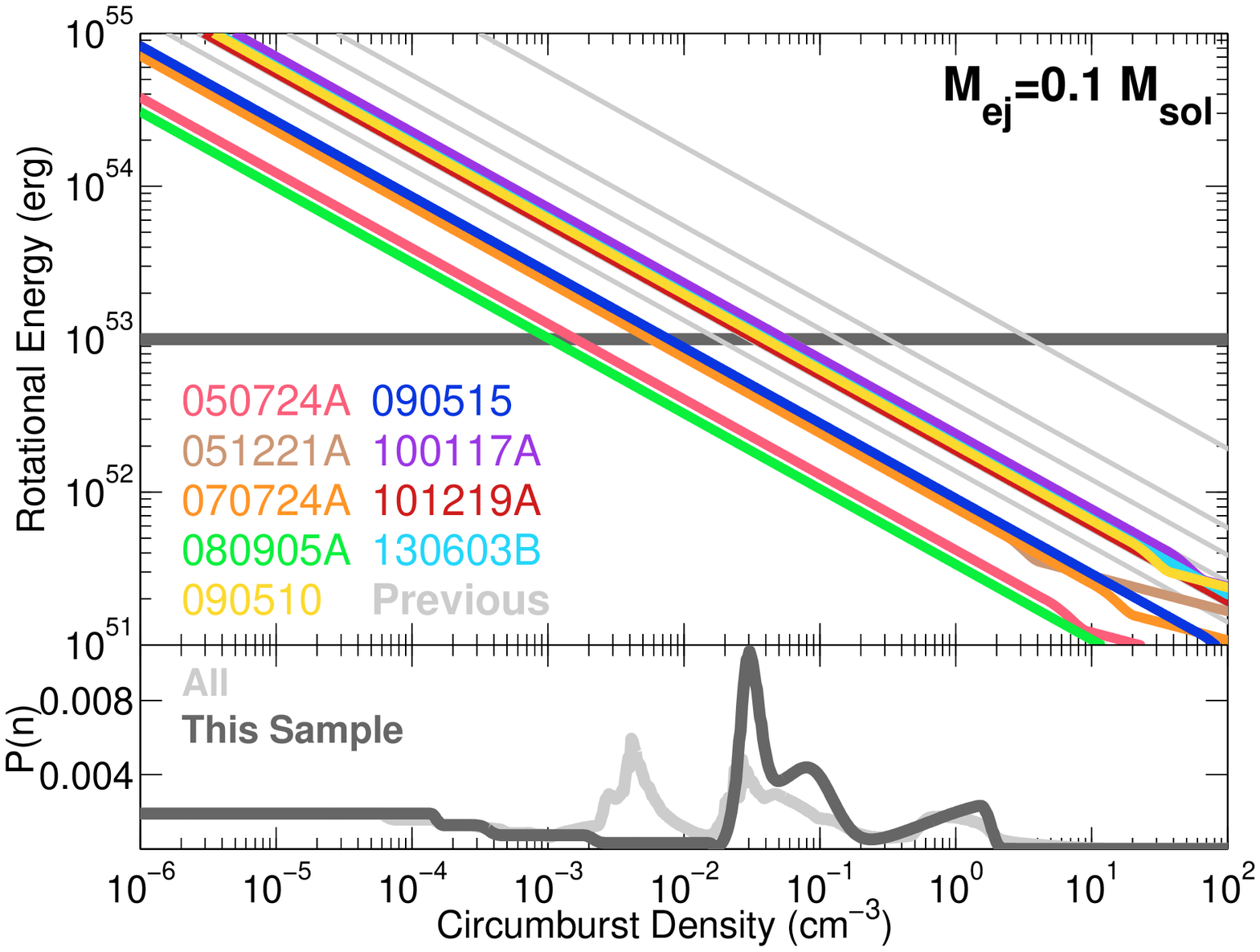}
\end{minipage}
\caption{Constraints on the rotational energy-circumburst density parameter space from VLA observations for the nine short GRBs in our sample (colored curves) for $M_{\rm ej}=0.03 ~M_{\odot}$ (left) and $M_{\rm ej}=0.1 ~M_{\odot}$ (right), assuming $\epsilon_e=\epsilon_B=0.1$ and $p=2.4$. The curves represent upper limits on the parameter space, where the region below each curve is allowed and the region above is ruled out.  Light grey curves denote constraints from previous work \citep{mb14,hhp+16}. A grey horizonal line represents the maximum extractable rotational energy of a $\sim 2.2 ~M_{\odot}$ magnetar of $10^{53}$~erg. In order to accommodate a magnetar with energy $10^{53}$~erg, the radio upper limits require that the circumburst densities are $\lesssim 4.8 \times 10^{-3}$~cm$^{-3}$ ($\lesssim 0.06$~cm$^{-3}$)  for ejecta masses of $0.03 ~M_{\odot}$ ($0.1 ~M_{\odot}$). Bottom panels show the distribution of densities as determined from afterglow observations for all short GRBs (light grey) and for the bursts in this sample (dark grey), where measurements have been weighted by their individual uncertainties.
\label{fig:En}}
\end{figure*}

To demonstrate the improvement upon previous samples, we compile the rotational energy-circumburst density constraints for the nine bursts in our sample for $\epsilon_B=0.1$ and the two ejecta masses, $M_{\rm ej}=0.03\,~M_{\odot}$ and $0.1\,~M_{\odot}$ (Figure~\ref{fig:En}), and compare these to the corresponding constraints from the two previous studies \citep{mb14,hhp+16}. We use the previously published radio limits, and assume the same values for $p$, $\epsilon_e$ and $\epsilon_B$ as in our study to ensure a uniform comparison. The resulting constraints are shown in Figure~\ref{fig:En}. The observations presented in this work provide deeper constraints on the combination of rotational energy and circumburst density by factors of $\approx 20-50$. For instance, for lower ejecta masses, previous works are able to rule out a magnetar with an available energy reservoir of $10^{53}$~erg for circumburst densities of $\gtrsim (1.5-250) \times 10^{-3}$~cm$^{-3}$, compared to $\gtrsim (0.09-4.8) \times 10^{-3}$~cm$^{-3}$ in this work (Figure~\ref{fig:En}). Also shown are the distributions of short GRB circumburst densities as inferred from their afterglows, for the nine bursts in this sample as well as the entire population. Overall, the large majority of short GRBs have low inferred densities of $\lesssim 0.1-1$~cm~$^{-3}$ (Figure~\ref{fig:En}; \citealt{fbm+15}). Thus, in comparison to previous studies, our work provides more meaningful limits in the density regime that actually corresponds to the inferred densities of short GRBs.

Incorporating these circumburst density measurements from the afterglows, we place limits on the maximum energy of a long-lived magnetar remnant of $\lesssim (0.02-17) \times 10^{53}$~erg for lower ejecta masses, and $\lesssim (0.05-55) \times 10^{53}$~erg for higher ejecta masses. Overall, our observations rule out a  magnetar energy of $10^{53}$~erg for half of the events in our sample.  Thus, we can rule out the presence of an {\it indefinitely} stable magnetar in a significant fraction of short GRBs with anomalous X-ray behavior.  However, we cannot rule out a relatively long-lived supramassive NS in these cases, which could survive for times approaching the magnetic dipole spin-down timescale \citep{spi06},

\begin{equation}
t_{\rm sd} \simeq 7 \left(\frac{B_{d}}{10^{15}\,{\rm~G}}\right)^{-2}\left(\frac{P}{{\rm1~ms}}\right)^{2}\,\,{\rm hr},
\label{eqn:tsd}
\end{equation}

\noindent where $P$ is the initial spin period and $B_{d}$ is the dipole surface magnetic field strength of the magnetar.
 
Our observations rule out a magnetar with an energy reservoir of $10^{52}$~erg associated with a single event, GRB\,050724A, which has a maximum allowed energy of $E_{\rm max} \approx (2-5) \times 10^{51}$~erg (Table~\ref{tab:param}). This event exhibited extended emission in the X-ray band which has previously been attributed to the spin-down energy of a long-lived magnetar \citep{gow+13}. If the merger that produced GRB\,050724A resulted in a remnant NS with a typical mass of $M_{\rm ns} \approx 2.3-2.4~M_{\odot}$ (e.g., \citealt{bok+08}, their Figure~4), this would imply that the NS EoS must be relatively soft at high densities, such that it supports a maximum (non-rotating) NS mass of $\lesssim 2.2~M_{\odot}$. This limit is also also supported by constraints from observations of radio pulsars (e.g., \citealt{of16}).  

If a remnant NS is responsible for the extended X-ray emission of GRB\,050724A and the NS EoS is indeed soft, the remnant NS could be on the high mass end of the supramassive range ($\gtrsim 2.5~M_{\odot}$), such that it imparts a relatively small amount of rotational energy of $\lesssim 5 \times 10^{51}$~erg to the surrounding medium before collapsing to a BH. In this case, the collapse time would be significantly less than the dipole spin-down timescale. By fitting the extended X-ray emission to a magnetar model, \citet{gow+13} derived an initial spin period of $P \approx 2.2$~ms and magnetic field strength of $B_d \approx 2.1 \times 10^{16}$~G, giving a dipole spin-down timescale of $t_{\rm sd} \approx 270$~sec (Equation~\ref{eqn:tsd}) which is indeed longer than the observed timescale of extended emission of $\approx 200$~sec. 

We note that the above conclusions are dependent on the value of $\epsilon_B$, and a very low value of $\epsilon_B$ would result in less stringent constraints on the energy of a magnetar. Our observational constraints would also be weakened if a large fraction of the magnetar rotational energy is emitted as gravitational waves instead of through electromagnetic spin-down (e.g., ~\citealt{dkp15,lg16,gzl16}).  However, this is unlikely during the supramassive NS phase unless the internal magnetic field is two orders of magnitude larger than the external dipole field (e.g., \citealt{dss+09}), or if the saturation amplitude of the $r$-mode instability is higher than commonly believed \citep{afm+03}.  

We compare our results to studies of magnetar emission from short GRBs at other wavelengths. \citet{gvo+15} used a broad-band afterglow model with time-varying energy injection due to spin-down of a long-lived magnetar to fit the X-ray and optical emission of four short GRBs with X-ray plateaus, and made predictions for the associated radio emission. For the two events which overlap with our sample, GRBs\,051221A and 130603B, the predicted radio emission is $\lesssim 1\,\mu$Jy at GHz frequencies at $\gtrsim 100$~days after the burst \citep{gvo+15}, consistent with the limits in this paper.  Similarly, fits to the luminosity and duration of short GRB X-ray plateaus with the magnetar model resulted in $\approx (0.01-6) \times 10^{52}$~erg of energy emitted during the plateau phase \citep{rom+13}, consistent with the values of $E_{\rm max}$ from our studies. In order to accommodate both the early-time X-ray activity and the limits on the long-term radio emission, we conclude that supramassive magnetars which inject a total energy of $\lesssim 10^{53}$~erg must be relatively common compared to stable magnetars. If the total energy constraints were uniformly more stringent, $\lesssim 10^{52}$~erg, the collapse to a BH should be relatively abrupt (i.e., during the plateau phase itself), and we would expect more events with dramatic drops in their X-ray light curves, similar to GRB\,090515 \citep{rot+10}.  

We note that the radio emission model (Section~\ref{sec:model}) applied in this paper neglects relativistic effects, both on the ejecta dynamics and the emission (e.g., relativistic Doppler beaming).  To understand how this simplification affects our results, we generate light curve models for the parameters considered by \citet{hhp+16}, who interpolate their results to also consider the relativistic limit.  We find that for models with ejecta mass of $0.1\,~M_{\odot}$, the peak fluxes and timescales are virtually identical across the full range of densities and energies considered here. However, for a lower ejecta mass of $0.01\,~M_{\odot}$, the peak fluxes are elevated by a factor of $\approx 10$ and the deceleration timescales are shortened by a factor of $\approx 2-3$ when compared to the Newtonian case. Therefore, incorporating relativistic effects only serves to make the predicted emission brighter, which then makes our observations even more constraining.

\section{Conclusions}
\label{sec:conc}

We study the long-term radio behavior of nine short GRBs with early-time excess emission in the X-ray band that may signify the presence of magnetars. Through our VLA observations on rest-frame timescales of $\approx 2-8$~yr after the bursts, we find no radio emission to luminosity limits of $\lesssim (0.05-8 )\times 10^{39}$~erg~s$^{-1}$ at $6.0$~GHz. Our study demonstrates that a significant fraction of short GRBs with anomalous X-ray behavior do not have the associated radio emission predicted from long-lived magnetars with energy reservoirs of $10^{53}$~erg. We also rule out a stable magnetar with an energy reservoir of $10^{52}$~erg in a single case, GRB\,050724A. These radio observations, together with the known X-ray behavior, imply that supramassive magnetars which inject $\lesssim 10^{53}$~erg of energy are common relative to stable magnetars.

Our study shows that a stiff NS EoS, corresponding to a maximum stable (non-rotating) NS mass of $M_{\rm ns} \gtrsim 2.3-2.4~M_{\odot}$, is disfavored, unless the NS mergers which give rise to short GRBs are particularly massive. However, population synthesis models suggest that such massive binaries only comprise a small fraction of all NS mergers \citep{bok+08}. A comparison of the observed rate of short GRBs to constraints on the NS merger rate from Advanced LIGO/Virgo will soon provide insight on the fraction of NS mergers which give rise to short GRBs, and thus additional insight on the NS EoS (e.g., \citealt{fbr+15}). Upcoming wide-field radio surveys will also constrain the population of long-lived magnetars \citep{mwb15}, independent of an association with short GRBs.

We cannot rule out that a long-lived magnetar is responsible for the extended X-ray activity after some short GRBs. However, we can conclude that most such remnants should be supramassive and hence should collapse to black holes on timescales which are comparable to or shorter than their magnetic dipole spin-down timescales. If future radio observations can uniformly constrain the total available energy from a magnetar to $\lesssim 10^{52}$~erg, we should expect more abrupt collapse signatures in the X-ray light curves of short GRB afterglows.  

The lack of evidence for stable, long-lived magnetars may impact observational signatures from NS mergers at other wavelengths. For example, neutron-rich outflows from the NS merger form heavy elements via the $r$-process and undergo radioactive decay, resulting in a kilonova transient \citep{lp98}. In the absence of a long-lived magnetar, the signal is expected to peak in the near-IR band on $\sim$week timescales due to the large opacities of the heavy elements produced \citep{bk13,kbb13,th13,gkr+14,ffh+15}. In contrast, the large neutrino luminosity from a long-lived magnetar may inhibit the formation of very heavy elements, resulting in a bluer kilonova which peaks at optical wavelengths on $\sim$day timescales \citep{mf14,kfm15}. If the sample in this paper is representative of all NS mergers, this supports the idea that kilonovae associated with NS mergers peak in the redder bands. 

Since NS mergers are expected to be strong sources of gravitational waves, similar searches for long-term radio emission following NS mergers detected within the Advanced LIGO/Virgo horizon distance of $200$~Mpc will be able to place limits of $\lesssim 6 \times 10^{36}$~erg~s$^{-1}$. Thus, such searches will be crucial in constraining the fraction of mergers that lead to magnetars with delayed or no collapse to a black hole, to significantly higher confidence than is possible with the cosmological sample. 

\begin{acknowledgments}
\noindent Support for this work was provided by NASA through Einstein Postdoctoral Fellowship grant number PF4-150121. BDM gratefully acknowledges support from NASA Fermi grant NNX14AQ68G, NSF grant AST-1410950, NASA ATP grant NNX16AB30G, and the Alfred P. Sloan Foundation. EB acknowledges support from NSF grant AST-1411763 and NASA  ADA grant NNX15AE50G. The National Radio Astronomy Observatory is a facility of the National Science Foundation operated under cooperative agreement by Associated Universities, Inc.
\end{acknowledgments}


\end{document}